\documentclass[12pt]{article}
\usepackage{epsfig}
\font\mybb=msbm10 at 12pt
\font\mybbsub=msbm10 at 10pt
\font\myeu=eufm10 at 12pt
\def\bb#1{\hbox{\mybb#1}}
\def\bbsub#1{\hbox{\mybbsub#1}}
\def\frak#1{\hbox{\myeu#1}}
\def\ZZ {\bb{Z}}
\def\NN {\bb{N}}
\def\ZZsub {\bbsub{Z}}
\def\RR {\bb{R}}
\def\CC {\bb{C}}
\def\g {\frak{g}}

\def\eps{\mbox{\bf e}}
\def\beq{\begin{equation}}
\def\eeq{\end{equation}}
\def\bdm{\begin{displaymath}}
\def\edm{\end{displaymath}}
\def\beqa{\begin{eqnarray}}
\def\eeqa{\end{eqnarray}}
\def\n{\nonumber\\}
\def\noi{\noindent}
\newcommand{\vs}[1]{\vspace{#1 mm}}

\begin{document}
\begin{titlepage}

\setcounter{page}{0}
\begin{flushright}
KEK Preprint 99-82 \\
AEI-1999-21 \\
hep-th/9909150
\end{flushright}

\vs{5}
\begin{center}
{\Large\bf On Discrete U-duality in M-theory\\}
\vs{10}

{\large
Shun'ya Mizoguchi\footnote{mizoguch@tanashi.kek.jp}} \\
\vs{5}
{\em Institute of Particle and Nuclear Studies \\
High Energy Accelerator Research Organization, KEK \\
Tanashi, Tokyo 188-8501, Japan} \\
\vs{5}
and \\
\vs{5}
{\large
Germar Schr\"oder\footnote{germar@aei-potsdam.mpg.de}
} \\
\vs{5}
{\it Max-Planck-Institut f\"ur Gravitationsphysik\\
Albert-Einstein-Institut\\
Am M\"uhlenberg 1, D-14424 Golm, Germany}\\
\end{center}
\vs{10}
\centerline{{\bf{Abstract}}}
\vskip 3mm
We give a complete set of generators for the discrete
exceptional U-duality groups of toroidal 
compactified type II theory and M-theory in $d\geq 3$. 
For this, we use the DSZ quantization in $d=4$ 
as originally proposed by Hull and Townsend, and determine 
the discrete group inducing integer shifts on the charge 
lattice. It is generated by fundamental unipotents, which 
are constructed by exponentiating the
Chevalley generators of the corresponding Lie algebra. 
We then extend a method suggested by the above authors 
and used by Sen for the heterotic string to get the 
discrete U-duality group in $d=3$, thereby obtaining a 
quantized symmetry in $d=3$ from a 
$d=4$ quantization condition.
This is studied first in a toy model, corresponding to 
$d=5$ simple supergravity, and then
applied to M-theory. It turns out that, in the toy model, the 
resulting U-duality group in $d=3$ is strictly smaller than the 
one generated by the fundamental unipotents corresponding to 
all Chevalley generators. However, for M-theory, both groups
agree.
We illustrate the compactification to $d=3$ by an embedding
of $d=4$ particle multiplets into the $d=3$ theory.

%\vskip 1cm\noindent
%{\it PACS} :  ??\\
%{\it Keywords} : ??

\end{titlepage}

\baselineskip=18pt
\setcounter{footnote}{0}

\section{Introduction and Motivation}

The study of nonperturbative duality symmetries 
in the last years has
dramatically changed our 
understanding of string theory.
The five perturbative string theories in ten dimensions 
are now understood as different limits
of a unified eleven dimensional theory, called M-theory, 
with a web of duality symmetries connecting them  
\cite{Tow95, Wit95}. Since these symmetries act 
on the string coupling constant, they are difficult to test. 
Nonetheless, several quantities protected by 
supersymmetry admit such tests, such as BPS masses and 
spectra as well as BPS saturated amplitudes.
Since the final form of a quantized M-theory is not yet
known, most 
tests use the low-energy effective limit, which is  
eleven dimensional supergravity. Global 
quantum symmetries in this limit should 
extend to a fully quantized theory. 

One of these nonperturbative dualities is the U-duality of
toroidially compactified type II string theory, 
introduced in \cite{Hul95}. It was conjectured 
that this symmetry is generated by the perturbative 
target space duality (for a review see \cite{Giv94}) and 
the strong-weak coupling duality \cite{Sdu} 
of the type II string 
\cite{Sch95} that do not commute, 
a phenomenon first discovered in
\cite{Sen95}.
For compactifications on the torus $\mbox{T}^d$ of type II 
string theory, 
the target space duality group corresponds to
$SO(d,d;\ZZ)$, while the S-duality 
of the type IIB theory acts as an $SL(2,\ZZ)$ group. 
The proposed U-duality group is then generated by

\beq
G(\ZZ)=SL(2,\ZZ) \bowtie SO(d,d;\ZZ),
\label{orig}
\eeq

\noi
where $\bowtie$ refers to the non-commuting 
action of both groups.
The type IIB S-duality was interpreted as the
modular group of the torus in the tenth and eleventh 
direction of M-theory \cite{Sch95,Asp96, Wit95}.
For compactifications of M-theory on $\mbox{T}^{d+1}$, 
the modular group
of the $d+1$-torus is $SL(d+1,\ZZ)$, containing the 
above $SL(2,\ZZ)$ 
as a subgroup. Therefore, rather than (\ref{orig}), 
the definition

\beq
G(\ZZ)=SL(d+1,\ZZ) \bowtie SO(d,d;\ZZ)
\label{orig2}
\eeq

\noi 
may be adopted.

The discussion of U-duality has led to 
identify nonperturbative states in string theory with 
higher dimensional branes wrapping on the internal torus
\cite{Hul95}, as well as to 
include D-branes into the theory
(see \cite{Pol96} for a review), improving dramatically 
our understanding of nonperturbative phenomena 
in string theory. 
Solitons of string theory corresponding to black hole 
solutions with regular horizon 
have made a microscopic interpretation
of  black hole entropy possible 
(see \cite{Maldacena} and \cite{Pee98} for a review). The 
entropy is known to be given by invariants of the U-duality 
group. 

In this context, the search 
for most general BPS black hole solutions 
has been important (\cite{Ber99}, see \cite{Dau99}, 
\cite{You97} for recent reviews).

U-duality, rephrased in an algebraic language \cite{Eli97}, 
has also been used in the infinite momentum frame of  
Matrix theory, predicting BPS states 
for compactifications not yet accessible to Matrix theory
and new ``mysterious'' states especially for compactifications 
to three dimensions. U-duality extends to a generalized 
electric-magnetic duality of the Super-YM theory. Upon 
inclusion of the lightlike momentum on the M-theory circle, 
the group was even proposed to be extended by rank,    
see \cite{Obe98} for a review 
and references therein. 

In \cite{Obe98}, a recent review was given on 
U-duality in diverse 
dimensions. The definition
(\ref{orig2}) was used for an algebraic definition 
of the U-duality group 
and was applied to study BPS 
spectra in diverse dimensions, to generate U-duality 
invariant mass formulae and to study $R^4$ corrections to 
type II theory as well as the implications on Matrix theory.

In this paper, we give generators for the discrete
U-duality in four dimensions,
thereby determining higher dimensional 
U-dualities as well, and apply these for a 
definition of U-duality in $d=3$. 
Rather than 
using the string inspired definition (\ref{orig2}) 
inferred into the low energy effective action, we 
will recall the original conjecture of
\cite{Hul95}, using the hidden symmetries of low energy 
supergravity 
\cite{Cre81}. 
The hidden symmetries have recently been 
reinvestigated in \cite{Cre98, Cre982}, 
where the symmetry groups $E_{d(+d)}$ were derived from 
$d=11$ diffeomorphisms in internal space, and the change 
of the symmetry group with respect to 
undualization of fields was investigated.

The low-energy effective
field theory of type II theory on $\mbox{T}^6$ is $d=4$ $N=8$ 
maximal supergravity and has
$E_{7(+7)}$ global symmetry \cite{Cre79} acting 
in the fundamental {\bf 56} representation. By an embedding 
into $E_{8(+8)}$, we will recover this representation in
an easily ``manageable'' way and give 
its Chevalley generators in this representation. This will 
allow to address the discrete group explicitly.

The 
Dirac-Schwinger-Zwanziger quantization condition in four
dimensions \cite{DSZ} discretizes electric and magnetic 
charges of the $U(1)$ gauge fields of the theory. 
The global $E_{7(+7)}$ symmetry is therefore broken to
a discrete $E_{7(+7)}(\ZZ)$ symmetry, inducing integer shifts
on the charge lattice. Since
the most general duality
group for our field configuration is known to be $Sp(2k,\RR)$
\cite{Gai81} and the discrete group $E_{7(+7)}(\ZZ)$ 
was not determined directly
in \cite{Hul95},  
it was conjectured that the group

\beq
E_{7(+7)}(\ZZ)= E_{7(+7)} \cap Sp(56,\ZZ)
\label{huto}
\eeq

\noi  
is a symmetry of the full type II string theory.

We will give a complete set of generators for the discrete 
$E_{7(+7)}(\ZZ)$ group by demanding integer 
shifts on the lattice. We will use these to identify the 
known string dualities in our notation, and show that the 
definitions (\ref{orig}) and (\ref{orig2}) agree with the one
we have found, 
and comment on the algebraic generators reviewed in 
\cite{Obe98}.  

Duality groups of the type II string theory in
dimensions larger than four are found by the intersections 
of the classical
duality group with
$E_{7(+7)}(\ZZ)$. 
For $d=3$, the situation is more complicated.

The classical continuous duality symmetry in three dimensions
is $E_{8(+8)}$, explicitly used for construction in 
\cite{Mar83}.
But the meaning of a duality group 
is not clear since only scalars remain in the
theory and electric charge seems ill defined. 
A DSZ quantization condition for $d=3$ seems unclear, 
and no analogue to $Sp(2k,\ZZ)$ and the construction 
(\ref{huto}) is known.

\begin{figure}[htbp]
\begin{center}
\leavevmode
\begin{picture}(0,0)%
\epsfig{file=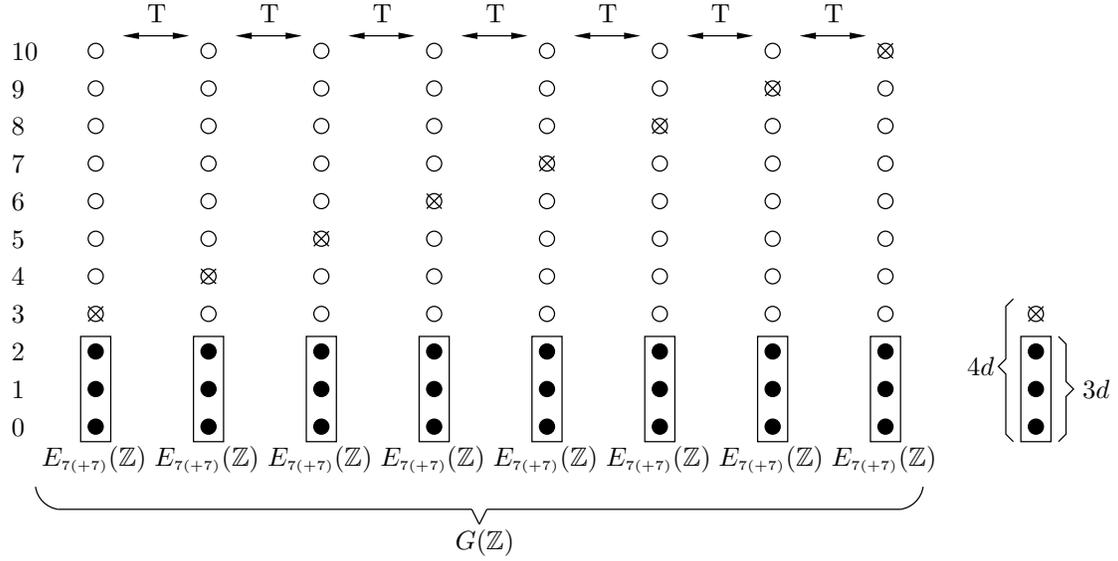}%
\end{picture}%
\setlength{\unitlength}{4144sp}%
\begin{picture}(6357,3270)(406,-2851)
\put(1216,344){\makebox(0,0)[lb]{\footnotesize T}}
\put(1891,344){\makebox(0,0)[lb]{\footnotesize T}}
\put(2566,344){\makebox(0,0)[lb]{\footnotesize T}}
\put(3241,344){\makebox(0,0)[lb]{\footnotesize T}}
\put(3916,344){\makebox(0,0)[lb]{\footnotesize T}}
\put(4591,344){\makebox(0,0)[lb]{\footnotesize T}}
\put(5266,344){\makebox(0,0)[lb]{\footnotesize T}}
\put(406,119){\makebox(0,0)[lb]{\footnotesize 10}}
\put(406,-106){\makebox(0,0)[lb]{\footnotesize 9}}
\put(406,-331){\makebox(0,0)[lb]{\footnotesize 8}}
\put(406,-556){\makebox(0,0)[lb]{\footnotesize 7}}
\put(406,-781){\makebox(0,0)[lb]{\footnotesize 6}}
\put(406,-1006){\makebox(0,0)[lb]{\footnotesize 5}}
\put(406,-1231){\makebox(0,0)[lb]{\footnotesize 4}}
\put(406,-1456){\makebox(0,0)[lb]{\footnotesize 3}}
\put(406,-1681){\makebox(0,0)[lb]{\footnotesize 2}}
\put(406,-1906){\makebox(0,0)[lb]{\footnotesize 1}}
\put(406,-2131){\makebox(0,0)[lb]{\footnotesize 0}}
\put(586,-2356){\makebox(0,0)[lb]
{\footnotesize  $E_{\scriptscriptstyle 7(+7)}(\ZZsub)$}}
\put(1261,-2356){\makebox(0,0)[lb]
{\footnotesize $E_{\scriptscriptstyle 7(+7)}(\ZZsub)$}}
\put(1936,-2356){\makebox(0,0)[lb]
{\footnotesize $E_{\scriptscriptstyle 7(+7)}(\ZZsub)$}}
\put(2611,-2356){\makebox(0,0)[lb]
{\footnotesize $E_{\scriptscriptstyle 7(+7)}(\ZZsub)$}}
\put(3286,-2356){\makebox(0,0)[lb]
{\footnotesize $E_{\scriptscriptstyle 7(+7)}(\ZZsub)$}}
\put(3961,-2356){\makebox(0,0)[lb]
{\footnotesize $E_{\scriptscriptstyle 7(+7)}(\ZZsub)$}}
\put(4636,-2356){\makebox(0,0)[lb]
{\footnotesize $E_{\scriptscriptstyle 7(+7)}(\ZZsub)$}}
\put(5311,-2356){\makebox(0,0)[lb]
{\footnotesize $E_{\scriptscriptstyle 7(+7)}(\ZZsub)$}}
\put(6811,-1906){\makebox(0,0)[lb]{\footnotesize $3d$}}
\put(6121,-1771){\makebox(0,0)[lb]{\footnotesize $4d$}}
\put(3061,-2851){\makebox(0,0)[lb]{\footnotesize $G(\ZZsub)$}}
\end{picture}
\end{center}
\caption{\label{fige8} Construction of three 
dimensional U-duality}
\end{figure}

We will give an explicit construction of the U-duality 
group in three
dimensions by extending a conjecture made by 
Hull and Townsend
\cite{Hul95}, parallel to a method applied to the heterotic
string by Sen \cite{Sen95, Sen953}.
This will enable us to describe the U-duality group in 
three dimensions
explicitly and give a set of generators for this group 
as well.
The
procedure is illustrated in figure \ref{fige8}. 
By compactifying
M-theory on the torus, we can choose eight 
different ways how to
compactify first to four dimensions. This
results in eight $E_{7(+7)}(\ZZ)$ acting {\it differently} 
on M-theory
fields. By reducing the theory further to three dimensions, 
these
groups are merged together to form the three dimensional 
duality group.

We will illustrate duality in $d=3$ by giving the embedding
of M-theory particle solutions in $d=4$, parallel to 
\cite{Sen95}.

\begin{figure}[htbp]
\begin{center}
\leavevmode
\begin{picture}(0,0)%
\epsfig{file=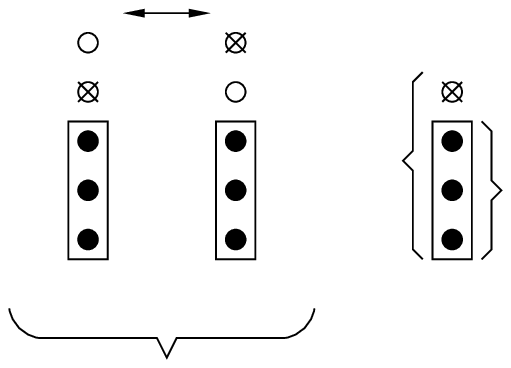}%
\end{picture}%
\setlength{\unitlength}{4144sp}%
\begin{picture}(2430,1875)(406,-2761)
\put(1216,-961){\makebox(0,0)[lb]{\footnotesize T}}
\put(406,-1231){\makebox(0,0)[lb]{\footnotesize 4}}
\put(406,-1456){\makebox(0,0)[lb]{\footnotesize 3}}
\put(406,-1681){\makebox(0,0)[lb]{\footnotesize 2}}
\put(406,-1906){\makebox(0,0)[lb]{\footnotesize 1}}
\put(406,-2131){\makebox(0,0)[lb]{\footnotesize 0}}
\put(2836,-1906){\makebox(0,0)[lb]{\footnotesize $3d$}}
\put(2146,-1771){\makebox(0,0)[lb]{\footnotesize $4d$}}
\put(1081,-2761){\makebox(0,0)[lb]{\footnotesize $G(\ZZsub)$}}
\put(616,-2356){\makebox(0,0)[lb]{\footnotesize $SL(2,\ZZsub)$}}
\put(1290,-2356){\makebox(0,0)[lb]{\footnotesize $SL(2,\ZZsub)$}}
\end{picture}
\end{center}
\caption{\label{figg2} Construction of three dimensional 
U-duality in the
$G_{2(+2)}$ toy model}
\end{figure}

Before turning to the $d=11$ case, we would
like to introduce the main concepts first in a 
toy model for simplicity.
For
this, we will use five dimensional simple supergravity, 
which upon
reduction to three dimensions exhibits a $G_{2(+2)}$ 
global symmetry
\cite{deW94, Miz98}. It is known that this theory closely 
resembles $d=11$
supergravity, the conjectured low energy limit of M-theory,
in many respects, but no string 
compactification described by this no-moduli supergravity at low 
energies is known \cite{Dab98}. 
The reduction procedure for this theory is
illustrated in figure \ref{figg2}. 

After introducing the 
main concepts in this simple model, the 
$d=11$ case will be constructed strictly along the 
same lines. Quite surprisingly, we will see that these models 
differ in a major point. We will discuss this 
difference again in the last section, where the direct 
embedding of the toy model into M-theory will be given.

\newpage

\section{The $G_{2(+2)}$ Toy Model}

We start with $d=5$ simple supergravity
\cite{Miz98, Cha80}. This theory is studied as a toy 
model for the low energy effective action of type II 
string theory and M-theory and introduce the main technical 
concepts. We will find an embedding into M-theory in 
the last section.

The bosonic part of the Lagrangian of simple
$d=5$ supergravity is given by

\bdm
{\cal L}
=- E^{(5)}(R^{(5)}+\frac14 F_{MN}F^{MN})
-\frac1{12\sqrt{3}}\epsilon^{MNPQR}F_{MN}F_{PQ}A_R ,
\edm

\noi
where $F_{MN}=2\partial_{[M}A_{N]}$. Indices $M,N$ 
run from $0...4$.
We take the signature $(+----)$.

To get a $d=3$ U-duality group, we are 
interested in compactifications of 
this theory to three 
dimensions.  
The vielbein can be
written as

\bdm
E_{~M}^{(5)A}=\left[
\begin{array}{cc}e^{-1} E_{~\mu}^{(3)\alpha}
& B_{\mu}^i e_i^a \\
0& e_i^a\end{array}
\right]
\edm
 
\noi
where the indices $i,a$ run from $1\dots 2$. $e_i^{\ a}$ 
is 
called the internal vielbein and is assumed to be of 
triangular 
shape. $e$ is its determinant. 
The vector field decomposes into
$A_M = (A_{\mu}, A_{2+i}),\ i=1\dots 2$, 
leaving a vector field and two scalars in three dimensions. 

In order to 
determine the $d=4$ U-duality analogue in this theory and
to see how it is embedded into $d=3$, 
we will need to 
reduce in a stepwise fashion. 

For this, we will 
interpret the coordinates $\{0,1,2,3\}$ as 
coordinates of the five dimensional theory 
with compact dimension 4, corresponding to the left side 
of figure \ref{figg2}. We will study 
this effectively four dimensional 
theory and
identify the U-duality group on the classical and 
quantum level.

We will then reduce the theory further to three dimensions
with coordinates $\{0,1,2\}$
and interpret the three dimensional 
theory as four dimensional theory with compact dimension 3. 
We will show that this reduction leads to a
$G_{2(+2)}/SO(4)$ nonlinear coset sigma model. We will identify 
the four dimensional U-duality in this theory and show how 
it acts.  

We will then study the
other compactification procedure indicated on the right side 
of figure \ref{figg2}. Now the coordinates $\{0,1,2,4\}$ will 
be interpreted as four dimensional coordinates, leading to a 
different U-duality group. We will then show how these two 
groups merge together to form the three dimensional 
U-duality.

\newpage
\subsection{The $d=4$ Theory}

Reducing the above Lagrangian to four dimensions yields
\cite{Miz98}

\beqa
{\cal L}
&=& - E^{(4)} R^{(4)}
+\frac32  E^{(4)} \partial_{\bar{\mu}}\ln\rho\; 
\partial^{\bar{\mu}}\ln\rho
+\frac12  E^{(4)} \rho^{-2}\; \partial_{\bar{\mu}}A_4\; 
\partial^{\bar{\mu}}A_4
\n
&&
-\frac14  E^{(4)} \rho^3 \;
B_{\bar{\mu}\bar{\nu}} B^{\bar{\mu}\bar{\nu}}
-\frac14  E^{(4)}  \rho\; F^{(4)}_{\bar{\mu}\bar{\nu}} 
F^{(4)\ \bar{\mu}\bar{\nu}}
\n
&&
-\frac1{4\sqrt{3}}
\epsilon^{\bar{\mu}\bar{\nu}\bar{\rho}\bar{\sigma}}
\left(A_4\;   F^{(4)}_{\bar{\mu}\bar{\nu}}
F^{(4)}_{\bar{\rho}\bar{\sigma}}
-A_4^2\; 
F^{(4)}_{\bar{\mu}\bar{\nu}}B_{\bar{\rho}\bar{\sigma}}
+\frac13 A_4^3\; B_{\bar{\mu}\bar{\nu}}B_{\bar{\rho}
\bar{\sigma}}\right)
\label{4dg2}
\eeqa

\noi
where the f\"unfbein and the vector field are parameterized as

\beq
E_{~M}^{(5)A}=\left[
\begin{array}{cc}\rho^{-\frac12}E_{~\bar{\mu}}^{(4)
\bar{\alpha}}
& \rho B_{\bar{\mu}} \\
0&\rho\end{array}
\right], \;
A_M = [A_{\bar{\mu}}, A_4].
\label{4dmetr}
\eeq

\noi
$\bar{\mu},\bar{\nu},\ldots$ are curved and $\bar{\alpha},
\bar{\beta},\ldots$
flat four dimensional indices. They run from $0\dots 3$.
We have defined $F^{(4)}_{\bar{\mu}\bar{\nu}} = 
F'_{\bar{\mu}\bar{\nu}} + B_{\bar{\mu}\bar{\nu}} A_4$.
$F'_{\bar{\mu}\bar{\nu}}
=2\partial_{[\bar{\mu}}A'_{\bar{\nu}]}$ is the field strength
of the Kaluza-Klein invariant vector field
$A'_{\bar{\mu}}=A_{\bar{\mu}}-B_{\bar{\mu}} A_4$, and 
$B_{\bar{\mu}\bar{\nu}} = 2 \partial_{[{\bar{\mu}}} 
B_{\bar{\nu}]}$.
$A'_{\hat{\mu}}$ is dualized in the standard way by adding a
Lagrange multiplier

\bdm
{\cal L}_{\rm Lag.mult.}=\frac12
\epsilon^{\bar{\mu}\bar{\nu}\bar{\rho}\bar{\sigma}}
\tilde{A}_{\bar{\sigma}}\partial_{\bar{\rho}}
F'_{\bar{\mu}\bar{\nu}}.
\edm

\noi
Defining the new field strength 
${\tilde A}_{\bar{\mu}\bar{\nu}}= 2 \partial_{[{\bar{\mu}}}
{\tilde A}_{\bar{\nu}]}$,
we introduce the vector notation

\bdm
{\cal G}_{\bar{\mu}\bar{\nu}}
=
\left[\begin{array}{c}\tilde{A}_{\bar{\mu}\bar{\nu}}\\
B_{\bar{\mu}\bar{\nu}}\end{array} \right],\
{\cal H}_{\bar{\mu}\bar{\nu}}=
\left[\begin{array}{c}H^{\tilde{A}}_{\bar{\mu}
\bar{\nu}}\\
H^B_{\bar{\mu}\bar{\nu}}\end{array} \right]
\edm

\noi
where

\bdm
H^{\tilde{A}}_{\bar{\mu}\bar{\nu}}=
-\frac2{E^{(4)}}\; \star \left(\frac{\delta{\cal L}}{\delta
{\tilde A}^{\bar{\mu}\bar{\nu}}}\right),\ \
H^B_{\bar{\mu}\bar{\nu}}=
-\frac2{E^{(4)}}\; \star \left(\frac{\delta{\cal L}}{\delta
B^{\bar{\mu}\bar{\nu}}}\right).
\edm

\noi
$\star\equiv\frac 12
E^{(4)-1}\epsilon^{\bar{\mu}\bar{\nu}\bar{\lambda}
\bar{\rho}}$
denotes the space-time dual in four dimensions.

The scalar fields of the theory are
$A_4$ and $\rho$. For this sector, we introduce a field
${\cal V}^{(4)}\in SL(2,\RR)/SO(2)$
in the
${\bf 4}$ irreducible representation. Using the Iwasawa 
decomposition,
${\cal V}^{(4)}$ is defined as

\bdm
{\cal V}^{(4)} =
P^{-1}\;\exp\left(-\frac 12 \ln \rho  H\right) 
\exp\left(-\frac1{\sqrt{3}} A_4 E\right)\; P
\edm

\noi
where the Chevalley generators $H,E,F$ of $SL(2,\RR)$ have 
been chosen
to be

\beq
H=\left[\begin{array}{cccc}3&&&\\&1&&\\&&-1&\\&&&-3
\end{array}\right],
E=\left[\begin{array}{cccc}
0&\sqrt{3}&&\\&0&2&\\&&0&\sqrt{3}\\&&&0\end{array}\right],
F=\left[\begin{array}{cccc}
0&&&\\\sqrt{3}&0&&\\&2&0&\\&&\sqrt{3}&0\end{array}\right]
\label{sl2alg}
\eeq

\noi
and

\beq
P = \left[\begin{array}{cccc}
0&0&0&1\\
1&0&0&0\\
0&0&1&0\\
0&1&0&0
\end{array}\right]=(P^{-1})^T.
\label{P}
\eeq

\noi
The explicit form of these generators will become 
important in the next section.

From ${\cal V}^{(4)}$ a field $P^{(4)}_\mu$ may be defined by

\bdm
\partial_{\mu}{\cal V}^{(4)} {\cal V}^{(4) -1}= 
Q^{(4)}_{\mu} + P^{(4)}_{\mu},
\ \ Q^{(4)}_{\mu} \in \frak{so}(2),
\ \ P^{(4)}_{\mu} \in \frak{sl}(2)-\frak{so}(2).
\edm

\noi
Using the above definitions, we note that the vector fields 
are related by
\bdm
{\cal F}_{\bar{\mu}\bar{\nu}} \equiv
\left[\begin{array}{c}
{\cal G}_{\bar{\mu}\bar{\nu}} \\ {\cal H}_{\bar{\mu}\bar{\nu}}
\end{array}\right]
=\Omega{\cal V}^{(4)T}{\cal V}^{(4)}
\left[\begin{array}{c}
\star {\cal G}_{\bar{\mu}\bar{\nu}} \\
\star {\cal H}_{\bar{\mu}\bar{\nu}}
\end{array}\right],
\edm

\noi
with the symplectic invariant

\bdm
\Omega=\left[\begin{array}{cc}&-{\bf 1}\\{\bf 1}&\end{array}
\right],
\edm

\noi
called twisted self duality in \cite{Cre98}, present in 
supergravities in all even dimensions.

Putting all together, the Lagrangian may 
be rewritten in the form
\cite{Miz98}

\bdm
{\cal L}
=  - E^{(4)} R^{(4)}+ 
E^{(4)} \left(\frac14 {\cal G}^T_{\bar{\mu}\bar{\nu}}
\star {\cal H}^{\bar{\mu}\bar{\nu}}
+ \frac3{10}{\rm Tr}(P^{(4)}_{\bar{\mu}}
P^{(4) \bar{\mu}})\right).
\edm

\subsection{Duality in $d=4$}

Following \cite{Hul95}, we will define the 
analogue of U-duality.
The classical theory has a continuous $SL(2,\RR)$ 
duality symmetry that interchanges Bianchi identities 
and equations of motion. Its action is

\beq
{\cal F}_{\hat{\mu}\hat{\nu}}\to\Lambda^{-1}
{\cal F}_{\hat{\mu}\hat{\nu}},\ \
{\cal V}^{(4)}\rightarrow {\cal V}^{(4)} \Lambda,
\ \ \Lambda \in SL(2,\RR).
\label{4dU}
\eeq

\noi
The transformation of ${\cal V}^{(4)}$ is accompanied by a 
compensating
local $SO(2)$ transformation
${\cal V}^{(4)}\rightarrow h(x){\cal V}^{(4)}\Lambda$,
$h(x)\in SO(2)$ to restore the parameterization
of the coset space.

If we define charges

\bdm
{\cal Z}= \left[\begin{array}{c}
p \\
q
\end{array}\right],\ \ \
p= \frac{1}{2\pi}\oint_\Sigma {\cal G}, \ \ \
q= \oint_\Sigma {\cal H}
\edm

\noi
and

\bdm
p=  \left[\begin{array}{c}
p^{\tilde{A}} \\
p^B
\end{array}\right],\ \ \
q=  \left[\begin{array}{c}
q_{\tilde{A}} \\
q_B
\end{array}\right],
\edm

\noi
the $p$ charges are interpreted as magnetic, 
the $q$ charges as
Noether electric charges.

We will introduce a charged 
``elementary'' soliton multiplet
in the $d=11$ section and study it. 
Due to its simplicity, the soliton solution presented 
there is actually a soliton 
in our toy-model as well. 
In M-theory, these solitons 
have been identified with fundamental and solitonic
string states and D-brane states that fill the 
U-duality multiplets \cite{Hul95}.

The charge vector ${\cal Z}$ transforms as

\bdm
{\cal Z} \to\Lambda^{-1}{\cal Z}, \ \Lambda \in SL(2,\RR)
\edm

\noi
under the classical duality symmetry.

Upon quantization, the charges
have to obey the Dirac-Schwinger-Zwanziger
charge quantization condition.
If all electric and magnetic charges exist,
they are restricted to live on an integer lattice,
and the $SL(2,\RR)$ symmetry is broken to a discrete subgroup 
inducing
integer shifts, which is $SL(2,\ZZ)$.
Note that the situation is complicated due to the 
fact that the
two gauge fields ${\tilde A}_{\bar{\mu}\bar{\nu}}$ and 
$B_{\bar{\mu}\bar{\nu}}$ are not treated on the same footing
by the $SL(2,\RR)$ symmetry, but $B_{\bar{\mu}\bar{\nu}}$ 
carries 
spin 3/2 and
${\tilde A}_{\bar{\mu}\bar{\nu}}$ spin -1/2 with respect to 
$SL(2,\RR)$, that is, they correspond to weights of different 
length.

To see how $SL(2,\ZZ)$ acts, we will not try to uncover the 
intersection
of $SL(2,\RR)$ with $Sp(4,\ZZ)$, but 
follow a more direct path. 
For this,
we change the basis of the $SL(2,\RR)$
representation space.
We define

\bdm
\tilde{\cal F}=U{\cal F}, \tilde{\cal V}^{(4)}=
U{\cal V}^{(4)}U^{-1},
\mbox{etc.}
\edm

\noi
with

\bdm
U=\left[\begin{array}{cccc}
1&&&\\&\sqrt{3}&&\\&&1&\\&&&\sqrt{3}\end{array}
\right].
\edm

\noi
Then the Lie algebra of $SL(2,{\bf R})$ is transformed into

\bdm
\tilde{E}=\left[\begin{array}{cccc}0&3&\\&0&2&\\&&0&1\\&&&0
\end{array}\right],
\tilde{F}=\left[\begin{array}{cccc}0&&&\\1&0&&\\&2&0&\\&&3&0
\end{array}\right],
\tilde{H}=H.
\edm

\noi
The DSZ quantization condition now reads
\beq
\tilde{\cal Z}^T\; U^{-1} \Omega\; U^{-1}\; \tilde{\cal Z}'
= \tilde{p}^{\tilde{A}} \tilde{q}_{\tilde{A}}' - 
\tilde{p}'^{\tilde{A}} \tilde{q}'_{\tilde{A}} 
+ \frac{1}{3}
(\tilde{p}^B \tilde{q}_B' - \tilde{p}'^B \tilde{q}'_B) = n,
\ \ n\in\ZZ.
\label{DSZshifted}
\eeq

The maximal subgroup of $SL(2,{\RR})$ preserving this 
discretization
is $SL(2,{\ZZ})$,
generated by
the modular group generators $P^{-1}SP$ and $P^{-1}TP$,
where
\bdm
S= \exp(-\tilde{F}) \exp \tilde{E} \exp(-\tilde{F})
=\left[\begin{array}{cccc}&&&1\\&&-1&\\&1&&\\-1&&&\end{array}
\right],
~~T= \exp \tilde{E}
=\left[\begin{array}{cccc}
1&3&3&1\\&1&2&1\\&&1&1\\&&&1\end{array}\right].
\edm

\noi
This symmetry can be interpreted as the analogue of $d=4$ 
U-duality in  our toy model.

If the $\{\tilde{p}^{\tilde{A}},\tilde{q}_{\tilde{A}}\} = 
\{p^A, q_A \} $ 
are chosen to be
integer, the $SL(2,\ZZ)$ symmetry and DSZ condition
yield that $\{\tilde{p}^B,\tilde{q}_B\}$ are in
$3\ZZ$.

Note that on the scalar $z\equiv -1/\sqrt{3} A_4 + i\rho$ 
this definition of
$SL(2,{\ZZ})$ induces the familiar modular 
transformations $z \to z+1$
and $z \to -1/z$ under $T$ and $S$,
respectively\footnote{Looking
at the Lagrangian, it is obvious that, in the asymptotic
limit, $z$ is exactly the $\tau$ parameter of 
electromagnetism plus theta
term for the field $F'_{\mu\nu}$, when $B_{\mu\nu}\equiv 0$. 
However, the above
$Sl(2,\ZZsub)$ will always mix all four types of charges 
and not 
preserve such a truncation.}.

We will now assume as in \cite{Sen95} 
that this quantum symmetry is not broken when 
we further compactify to three dimensions. 

To recover $d=4$ solitons in the $d=3$ theory,
one may take an array of solitonic solutions 
aligned along a specific direction 
\cite{Sen95, Hul95, Maldacena, Obe98}. 
Compactification of this direction then corresponds
to identify this array periodically and taking the period 
to be small. We will see in the M-theory section that
this leads to vortex solutions in three dimensions. We 
will recover the $d=4$ 
U-duality in the reduced theory acting exactly on these 
solutions and their $d=4$ charges. 
The equivalence of such periodic 
array solutions and fundamental string states under duality
for the
$d=3$ heterotic string 
was studied closely in \cite{Sen95}. 

\subsection{The $d=3$ Theory}

We will now reduce the theory further to three dimension
\footnote{The reduction from $d=4$ to $d=3$ in the context of 
black holes in Kaluza Klein theory 
was studied in \cite{Brei88}.}.
For this, we assume that the direction 3 is compact and 
the fields do not depend on this coordinate, which 
corresponds to keeping the zeroth Fourier component with 
respect to the compact direction. 

We will ``maximally dualize'' \cite{Cre98} all fields
such that only 
scalars remain in the theory.
Quite analogously to the reduction step before, we choose the 
vierbein to be
\bdm
E_{~\bar{\mu}}^{(4)\bar{\alpha}}=\left[
\begin{array}{cc}e^{\phi/2} E_{~\mu}^{(3){\alpha}}
& e^{-\phi/2} \hat{B}_{\mu} \\
0&e^{-\phi/2}\end{array}
\right],
\edm     

\noi
where $\mu,\alpha$ now run from $0\dots 2$
and define the $U(1)$ vector field strengths
${\cal G}'_{\mu\nu}=2\partial_{[\mu}{\cal G}'_{\nu]}$ by 
taking
${\cal G}'_{\mu}={\cal G}_{\mu}-\hat{B}_{\mu} {\cal G}_3$.

The vector fields are then dualized by adding the Lagrange 
multiplier

\bdm
{\cal L}_{\rm Lag.mult.}=\frac12
\epsilon^{{\mu}{\nu}{\rho}}
\bar{\eta} \partial_{{\rho}}{\cal G}'_{{\mu}{\nu}}
\edm

\noi
and integrating out ${\cal G}'_{{\mu}{\nu}}$. This yields

\beq
\partial_{\mu} \bar{\eta} = {\cal H}_{\mu 3}
\label{4dH}
\eeq

\noi
which together with the definition

\bdm
\partial_{\mu} \eta = {\cal G}_{\mu 3}
\edm

\noi
and

\bdm
\cal{Y}=  \left[\begin{array}{c}
\eta \\
\bar{\eta}
\end{array}\right]
\edm

\noi
will be of crucial importance in the following.

It remains to dualize the Kaluza-Klein field strength 
$\hat{B}_{\mu \nu}$ by
adding

\bdm
{\cal L}_{\rm Lag.mult.}=\frac{1}{2}
\epsilon^{{\mu}{\nu}{\rho}}
f \partial_{{\rho}}\hat{B}_{{\mu}{\nu}}
\edm

\noi
which by integrating out $\hat{B}_{{\mu}{\nu}}$  yields

\bdm
\partial_{\mu} f = - \frac12 E^{(3)} \epsilon^{{\mu}{\nu}{\rho}} 
e^{-2\phi}
 \hat{B}_{{\nu}{\rho}} -  \frac12 {\cal Y}^t \Omega 
\partial_{\mu} 
{\cal Y}.
\edm

\noi
The Lagrangian becomes

\beqa
{\cal L}
&=& - E^{(3)} R^{(3)}
+\frac12  E^{(3)} \partial_{\mu}\phi\; \partial^{\mu}\phi
+\frac{3}{10}
E^{(3)}{\rm Tr}(P^{(4)}_{\mu}P^{(4) \mu})
\n
&&
+\  E^{(3)} e^{2\phi} (\partial_{\mu} f + {\cal Y}^t 
\Omega \partial_{\mu}
 {\cal Y})
 (\partial^{\mu} f + {\cal Y}^t \Omega \partial^{\mu} {\cal Y})
\n
&&
+ 2 E^{(3)} e^{\phi} \partial_{\mu} {\cal Y}^t
{\cal V}^{(4) t} {\cal V}^{(4)} \partial^{\mu} {\cal Y}.
\label{g2}
\eeqa

\subsection{The $G_{2(+2)}$ Coset in $d=3$}

We will now show that the scalars of this theory fit into a
$G_{2(+2)}/SO(4)$ coset, where $G_{2(+2)}$ is the 
normal real form \cite{Gil74} of $G_2$. The Lagrangian gets

\beq
{\cal L}
= - E^{(3)} R^{(3)}
+\frac{1}{16}
E^{(3)}{\rm Tr}(P^{(3)}_{\mu}P^{(3) \mu})
\label{g2coset}
\eeq

\noi
where the scalars are contained in a field ${\cal V}^{(3)}\in
G_{2(+2)}/SO(4)$ and we define

\bdm
\partial_{\mu}{\cal V}^{(3)} {\cal V}^{(3) -1}= 
Q^{(3)}_{\mu} + P^{(3)}_{\mu},
\ \ Q^{(3)}_{\mu} \in \frak{so}(4),
\ \ P^{(3)}_{\mu} \in \frak{g}_{2(+2)}-\frak{so}(4).
\edm 

For the algebra $\frak{g}_{2(+2)}$ of $G_{2(+2)}$,
we will use 
Freudenthal's realization of exceptional Lie algebras
explained in appendix A. This will allow us to stay as 
closely as possible
to the $d=11$ case. 

For the positive roots,
the corresponding generators are
$E^i_{\ j}, 1\le i\le j \le 3$, $E^i, 1\le i\le 2$ and $E^*_3$, 
for the Cartan subalgebra we use generators $h_i,\ i=1,2$. 
Their definitions and commutators are given in the appendix.
The generator $E^1_{\ 2}$ corresponds to the long simple 
positive root,
the generator $E^2$ to the short one.

We will now use the fact that $G_{2(+2)}$ has a maximal subgroup
$SL(2)\times SL(2)$, where the two $SL(2)$ groups are 
generated by the
short simple and the lowest root. In the Iwasawa decomposition,
the $SL(2)$ generated by the short simple root will be 
associated with
${\cal V}^{(4)}$, while the $SL(2)$ generated by the lowest root
will carry the $3d$ dilaton and dualized
Kaluza-Klein gauge field.

\begin{figure}[htbp]
\begin{center}
\leavevmode
\begin{picture}(0,0)%
\epsfig{file=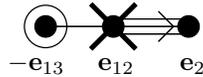}%
\end{picture}%
\setlength{\unitlength}{4144sp}%
\begin{picture}(1152,408)(1621,-592)
\put(2651,-556){\makebox(0,0)[lb]
{\footnotesize $\eps_2$}}
\put(2161,-556){\makebox(0,0)[lb]
{\footnotesize $\eps_{12}$}}
\put(1621,-556){\makebox(0,0)[lb]
{\footnotesize $-\eps_{13}$}}
\end{picture}
\end{center}
\caption{\label{g2decomp} Decomposition of 
$G_{2(+2)}\supset SL(2)\times SL(2)$. The root surrounded by a 
circle is the lowest root added to the Dynkin diagram.}
\end{figure}

\noi
To see this, we define the vector

\bdm
{\cal S}^t= (\frac1{\sqrt{3}} E^*_3, -E^1_{\ 2} , 
\frac1{\sqrt{3}} E^1,
E^2_{\ 3}).
\edm  

\noi
From the $\frak{g}_{2(+2)}$
commutation relations given in the appendix
one may then easily deduce that

\bdm
[E^2,{\cal S}_i] = (P^{-1} E P)^t_{ij}\ {\cal S}_j, \ \ \
[E^*_2,{\cal S}_i] = (P^{-1} F P)^t_{ij}\ {\cal S}_j,
\edm

\noi
with

\bdm
[{\cal S}_i, {\cal S}_j] = \Omega_{ij} E^1_{\ 3}
\edm

\noi
where the $E,F$ are the $\frak{sl}(2,\RR)$ generators given
in (\ref{sl2alg}) and $P$ is given in (\ref{P}).  

Using these relations, it is straightforward to verify that the 
Lagrangian
(\ref{g2}) is identical to (\ref{g2coset}) 
if\footnote{Note that {\it no} representation of $G_{2(+2)}$ 
needs to be specified for this. The calculation uses the 
algebra, the representation influences 
only the prefactor of the scalar term in the Lagrangian which 
could be chosen according to which representation is used. 
But due to the decomposition (\ref{g2dec}), we have 
used the adjoint of $G_{2(+2)}$.}  

\beqa
{\cal V}^{(3)} &=& \exp\left({-\frac12\ln\rho\ (h_2-h_1)}\right)
\exp\left({-\frac1{\sqrt{3}} A_4\ E^2}\right)
\n
&&
\exp\left(\frac12\phi\ (h_1+h_2)\right)
\exp\left({{\cal Y}_i\ {\cal S}_i}\right)
\exp\left({f\ E^1_{\ 3}}\right)
\n
&=&
{\cal V}'^{(4)}
\exp\left(\frac12\phi\ (h_1+h_2) \right)
\exp\left({{\cal Y}_i\ {\cal S}_i}\right)
\exp\left({f\ E^1_{\ 3}}\right).
\label{3dg21}
\eeqa

\noi
The prime refers to the fact that ${\cal V}'^{(4)}$ is not 
in the spin 3/2 representation introduced in (\ref{sl2alg}) 
of $SL(2,\RR)$ any more.
$G_{2(+2)}$ decomposes as

\beqa
G_{2(+2)}&\supset& SL(2,\RR)\times SL(2,\RR)
\n
{\bf 14}&\supset&({\bf 2},{\bf 4}) + ({\bf 1},{\bf 3}) + 
({\bf 3},{\bf 1})
\label{g2dec}
\eeqa

\noi
In the {\bf 14} representation, ${\cal V}'^{(4)}$, corresponding to
the latter $SL(2,\RR)$, 
therefore has block structure

\bdm
{\cal V}'^{(4)}=\left[\begin{array}{cccccc}
\fbox{\bf 4}&&&&&\\&\fbox{$\bar
{\bf 4}$}&&&&\\&&\fbox{\bf 3}&&&\\&&&
\fbox{\bf 1}&&\\&&&&\fbox{\bf 1}&
\\&&&&&\fbox{\bf 1}\end{array}\right],
\edm

\noi
containing the familiar {\bf 4} block of $d=4$.

\subsection{Identifying $d=4$ U-Duality in the $d=3$ Theory}

Defining an element

\bdm
\Lambda\in SL(2,\RR), \ \ \
\Lambda = e^X, \ X= a\; E^2 + b\; E^*_2 + c\; (h_2 -h_1), 
\ a,b,c\in \RR
\edm

\noi
we may look at

\beqa
{\cal V}^{(3)}\rightarrow{\cal V}^{(3)} \Lambda^{-1}
&=&
{\cal V}'^{(4)} \Lambda^{-1}
\exp\left({\frac12\phi\ (h_1+h_2)}\right) \Lambda 
\exp\left({{\cal Y}_i\  
{\cal S}_i}\right) \Lambda^{-1}
\exp\left({f\ E^1_{\ 3}}\right)
\n
&=&
{\cal V}'^{(4)} \Lambda^{-1}
\exp\left({\frac12\phi\ (h_1+h_2) }\right) 
\exp\left({{\cal Y}_i
[\exp D(X)]^t_{ij}\ {\cal S}_j}\right)
\exp\left({f\ E^1_{\ 3}}\right)
\n
&=&
{\cal V}'^{(4)} \Lambda^{-1}
\exp\left({\frac12\phi\ (h_1+h_2)}\right) \exp\left({
[D(\Lambda) {\cal Y}]_i\ {\cal S}_i}\right)
\exp\left({f\ E^1_{\ 3}}\right)
\label{recover}
\eeqa

\noi
therefore

\bdm
{\cal V}'^{(4)}\rightarrow {\cal V}'^{(4)} \Lambda^{-1} 
\ \mbox{and}\
{\cal Y}\rightarrow D(\Lambda) {\cal Y}
\edm

\noi
where $D(.)$ is the spin 3/2 representation 
introduced in (\ref{sl2alg}).
Since the vector ${\cal Y}$ carries the $d=4$  
charges,
this is exactly the transformation behavior (\ref{4dU}) 
and resembles the
four dimensional U-duality in the reduced model. 
We will 
illustrate this in the $d=11$ case by studying an 
``elementary'' soliton multiplet. 

In the reduced model,
the discrete group is now generated by

\bdm
S^2=\exp(-E^*_2)\exp(E^2)\exp(-E^*_2), \ T^2=\exp(E^2).
\edm

\subsection{Connection to $d=5$ Fields}

We will now take a step back and look at the reduction 
we performed so far. The choice we took for the $d=5$
vielbein was
\beq
E_{~M}^{(5)A}=
\left[
\begin{array}{cc}e^{-1}E_{~\mu}^{(3)\alpha}
& B_{\mu}^i e_i^a \\
\mbox{\Large 0}
&\makebox[3.7cm]
{\rule[-0.7cm]{0cm}{1.5cm}$e_i^a$}
\end{array}
\right]
=
\left[
\begin{array}{cc}
e^{\phi/2}\rho^{-\frac12}E_{~\mu}^{(3)\alpha}
& 
\begin{array}{cc}
e^{-\phi/2}\rho^{-\frac12}\hat{B}_{\mu} & \rho B_{\mu}
\end{array} \\
\mbox{\Large 0}
&\parbox{3.7cm}{$
\rule[-0.5cm]{0cm}{1.3cm}
\begin{array}{cc}
\rho^{-\frac12}e^{-\phi/2} & \ \ \ \ \rho B_3 \\
 0 & \rho \end{array}$}
\end{array}
\right].
\label{fuenfbein}
\eeq

\noi
In order to compare the different compactification in figure 
\ref{figg2} and join the $d=4$ U-duality groups together
to a $d=3$ U-duality group, we reexpress the 
coset matrix (\ref{3dg21}). 

We define 

\beqa
\varphi &=& \eta^{\tilde{A}} + \frac{1}{\sqrt{3}} A_3 A_4 -
\frac{1}{\sqrt{3}} B_3 A_4^2,
\n
\Psi_1 &=& f + \frac12 B_3 \Psi_2 - \frac14 B_3 A_4 \varphi +
\frac{1}{6\sqrt{3}} A_3 A_4 (A_3 - B_3 A_4),
\n
\Psi_2 &=& \bar{\eta}_B -\frac12 A_4 \varphi - 
\frac{1}{3\sqrt{3}} A_4^2 (A_3 -B_3 A_4).
\nonumber
\eeqa

\noi
Note that the new field are polynomial in the old fields. 
One may now use (\ref{fuenfbein}) and (\ref{G2generatorrelations}) to get

\beqa
{\cal V}^{(3)} &=& 
\exp\left(-\ln(e_1^{\ \dot{1}})\ h_1 
-\ln(e_1^{\ \dot{1}}e_2^{\ \dot{2}})\ h_2\right)
\exp \left(-e_1^{\ \dot{2}}e_{\dot{2}}^{\ 2}\ E^1_{\ 2}\right) 
\n
&&
\exp\left(\Psi_i\ 
E^i_{\ 3}\right)
\n
&&
\exp \left(\frac{1}{\sqrt{3}} 
\left(-A_{2+i}\ E^i + \varphi\ E^*_3 
\right)\right)
\n
&=& \exp\left(\frac12\left(
(\phi + \ln\rho)\ h_1 + (\phi -\ln\rho)\ h_2\right)\right)
\n
&&\exp \left(-B_3\ E^1_{\ 2}\right) 
\exp\left(\Psi_1\ E^1_{\ 3}+\Psi_2\ E^2_{\ 3}\right)
\n
&&
\exp \left(\frac{1}{\sqrt{3}} 
\left(-A_{3}\ E^1  -A_{4}\ E^2  + \varphi\ E^*_3 
\right)\right)
\label{one}
\eeqa

\noi
where dotted indices are flat internal indices. 
Explicitly, $\varphi$ and $\Psi_{i}$ obey

\beqa
\partial_{\mu} \varphi &=& 
- e^2 E^{(3)} \epsilon_{\mu\nu\rho} 
(\partial^{[\nu}A^{\rho]} + B^{i [\nu}\partial^{\rho]}A_{(2+i)})
+\frac{1}{\sqrt{3}}\epsilon^{ij} A_{2+i} \partial_{\mu} A_{2+j}, 
\n
\partial_{\mu} \Psi_i &=& 
-\frac12 e^2 E^{(3)} \epsilon_{\mu\nu\rho} B^{\nu\rho}_i 
-\frac12(\varphi 
\partial_{\mu} A_{2+i} - \partial_{\mu}\varphi A_{2+i}) -
\frac{1}{3\sqrt{3}}\epsilon^{jk} A_{2+i} A_{2+j} 
\partial_{\mu} A_{2+k}.
\nonumber
\eeqa

\noi
This exactly reproduces the result of \cite{Miz98} obtained 
by direct reduction to $d=3$.

\subsection{Different Orders of Compactification} 

We will now exploit this connection to $d=5$ fields.
We chose to reduce in 
the order of the left side of figure \ref{figg2} and arrived at
a theory with scalar coset matrix (\ref{one}) called 
${\cal V}^{(3)}_{\# 1}$ in the following. The $d=4$ U-duality
is generated by 

\bdm
S^2=\exp(-E^*_2)\exp(E^2)\exp(-E^*_2), \ T^2=\exp(E^2).
\edm
  
We will now turn to the right side of figure \ref{figg2}.
We consider first the $d=5$ vielbein.
We need a convenient parameterization for the second reduction.
For this, we perform a local Lorentz transformation 

\bdm
\Lambda_B^{\ A}=\left[
\begin{array}{cc}
\delta^{\ \alpha}_{\beta}
& 0 \\
\mbox{\Large 0}
&\begin{array}{cc}
\cos{\theta}& -\sin{\theta}  \\
 \sin{\theta}&\ \  \cos{\theta}
\end{array}
\end{array}
\right]
\edm

\noi 
such that 

\bdm
\left[
\begin{array}{cc}
\rho' & 0 \\
\rho' B_3' & \rho'^{-\frac12}e^{-\phi'/2} \end{array}
\right]
=
\left[
\begin{array}{cc}
\rho^{-\frac12}e^{-\phi/2} & \ \ \ \ \rho B_3 \\
 0 & \rho \end{array}
\right]
\left[
\begin{array}{cc}
\cos{\theta}& -\sin{\theta}  \\
 \sin{\theta}& \cos{\theta}
\end{array}
\right]  
\edm

\noi
which yields $\tan{\theta}=\rho^{3/2}e^{\phi/2} B_3$.

Using this parameterization, we can perform the reduction 
strictly parallel to the one before, only a sign change in the
Chern-Simons term has to be taken into account.
The result is

\beqa
{\cal V}^{(3)}_{\# 2} &=& 
\exp\left(\frac12\left(
(\phi' + \ln\rho')\ h_1 + (\phi' -\ln\rho')\ h_2\right)\right)
\n
&&\exp \left(-B'_3\ E^1_{\ 2}\right) 
\exp\left(\Psi_2\ E^1_{\ 3}+ \Psi_1\ E^2_{\ 3}\right)
\n
&&
\exp \left(\frac{1}{\sqrt{3}} 
\left(A_{4}\ E^1 + A_{3}\ E^2 - \varphi\ E^*_3 
\right)\right).
\nonumber
\eeqa

\noi
The $d=4$ discrete U-duality
is again generated by 

\bdm
S^2=\exp(-E^*_2)\exp(E^2)\exp(-E^*_2), \ T^2=\exp(E^2).
\edm

\subsection{Joining $d=4$ U-dualities in $d=3$}

In order to join the two compactifications and U-dualities,
we recognize that the two scalar matrices are related by

\beq
{\cal V}^{(3)}_{\# 2} = 
(P S^1_{\ 2})^{-1} \; \exp\left(\theta(E^1_{\ 2} - E^2_{\ 1})
\right)\;\ 
{\cal V}^{(3)}_{\# 1}\;\ P S^1_{\ 2} 
\label{relation}
\eeq

\noi
$\theta$ is given by $\tan{\theta}=\rho^{3/2}e^{\phi/2} B_3$ 
as above, the factor $\exp(\theta(E^1_{\ 2} - E^2_{\ 1}))
\in SO(4)$ is
nothing but the local Lorentz transformation we performed.
$S^1_{\ 2}$ is given by

\bdm
S^1_{\ 2}=\exp(-E^2_{\ 1})\exp(E^1_{\ 2})\exp(-E^2_{\ 1})
\edm

\noi 
and corresponds to the T-duality transformation in figure 
\ref{figg2}. It represents the needed Weyl reflection in the 
root space of $\frak{g}_{2(+2)}$.

Somewhat unexpected is the appearance of $P$ in 
(\ref{relation}). 
$P$ is a ``parity'' transformation given by 

\bdm
P= (-1)^{(h_1+h_2)} = (S^1_{\ 3})^2 = (S^2)^2
\edm 

\noi
and is an element of $d=4$ U-duality corresponding 
to a charge conjugation of the $d=4$ charges. 

Turning to the U-duality transformations, consider a 
transformation $U$ on ${\cal V}^{(3)}_{\# 1}$.
We have

\beqa
{\cal V}'^{(3)}_{\# 2}&=&
(P S^1_{\ 2})^{-1}\;\exp\left(\theta(E^1_{\ 2} - E^2_{\ 1})
\right)\ {\cal V}'^{(3)}_{\# 1}\;\ P S^1_{\ 2}
\n
&=& (P S^1_{\ 2})^{-1}\;\exp\left(\theta(E^1_{\ 2} - E^2_{\ 1})
\right)\ {\cal V}^{(3)}_{\# 1}\;\ P S^1_{\ 2}
\; \ (P S^1_{\ 2})^{-1}\;\ U\ (P S^1_{\ 2})
\n
&=&
{\cal V}^{(3)}_{\# 2} 
\; \ (P S^1_{\ 2})^{-1}\;\ U\ (P S^1_{\ 2})
\n
&=&
{\cal V}^{(3)}_{\# 2}\;\ \tilde{U}
\nonumber
\eeqa

\noi
$\tilde{U}$ is a matrix generated by

\bdm
S^1=\exp(-E^*_1)\exp(E^1)\exp(-E^*_1), \ T^1=\exp(E^1).
\edm

\noi 
The joint U-duality group $U(\ZZ)$ 
in three dimensions is therefore generated by  

\bdm
S^1,\ S^2,\ T^1,\ T^2.
\edm

This discrete subgroup of $G_{2(+2)}$ can now be compared 
with 
the notion of $G_{2(+2)}(\ZZ)$ as presented in appendix B. 
We will see there that
$G_{2(+2)}(\ZZ)$ is expected to be generated by  
$S^i_{\ j},\ S^i,\ T^i_{\ j},\ T^i$ for all positive roots of 
$\frak{g}_{2(+2)}$. 

But quite obviously $U(\ZZ)$ is {\it not}
$G_{2(+2)}(\ZZ)$, but the former is strictly {\it smaller}
than the latter. All generators of $U(\ZZ)$ correspond to 
short roots of $\frak{g}_{2(+2)}$. It is only possible to 
generate e.g. $(T^1_{\ 2})^3$ from them, but not $(T^1_{\ 2})$.  
That these groups do not agree is therefore connected to the 
fact that $G_{2(+2)}$ is not simply laced. 

$d=5$ simple supergravity has no moduli fields, and no string 
compactification described by this supergravity at low energies 
is known. Therefore one cannot determine which is the 
``correct'' 
U-duality group until such a microscopic realization is found.

Note that the proof in the appendix and in \cite{Matsumoto} 
uses minimal resp. basic representations, that is, all 
nontrivial weights of the representation are transformed into 
each other by the Weyl group and must therefore be of the 
same length. The adjoint of $G_{2(+2)}$ and the {\bf 4} of 
$SL(2,\RR)$ do not belong to this class. 
But, as stated in \cite{Matsumoto},  
the notion of the group $G_{2(+2)}(\ZZ)$ is 
independent of its representation.

The disagreement of $U(\ZZ)$ with $G_{2(+2)}(\ZZ)$ might 
be surprising at first. We will now turn to the $d=11$ case to
see if the same is true in this model.

\section{The $d=11$ Case}

M-theory is supposed to have eleven dimensional supergravity 
as low energy limit. The bosonic part of the 
Lagrangian is given by

\beq
{\cal L}
=-\frac14 E^{(11)}(R^{(11)}+\frac1{12} F_{MNPQ}F^{MNPQ})
+\frac2{12^4}\epsilon^{MNPQRSTUVWX}F_{MNPQ}F_{RSTU}A_{VWX},
\label{11d}
\eeq

\noi
where $F_{MNPQ}=4\partial_{[M}A_{NPQ]}$. Indices $M,N,\dots$ 
now run from $0...10$, the metric has signature $(+--\dots -)$.

Quite analogously to the last section, 
if the theory is compactified to three dimensions, 
the vielbein can taken to be of the form

\bdm
E_{~M}^{(11)A}=\left[
\begin{array}{cc}e^{-1} E_{~\mu}^{(3)\alpha}
& B_{\mu}^{(3) i} e_i^{\ a} \\
0& e_i^{\ a}\end{array}
\right].
\edm
 
Again, we define the internal vielbein $e_i^{\ a}$ 
with indices $i,a$ running in $1\dots 8$. $e_i^{\ a}$ is 
assumed to be of triangular shape in the following.

In order to identify the $d=4$ U-duality in $d=3$, 
we will proceed exactly as in the toy model in a 
stepwise fashion. 

We will first study M-theory compactified to $d=4$ and 
U-duality in $d=4$.

\subsection{\label{chap} The $d=4$ Theory}

The reduction to $d=4$ was carried out in detail e.g. 
in \cite{Cre79} 
(see also \cite{Cre98} for a more recent treatment).   
The reduced Lagrangian reads

\beqa
{\cal L}
&=& -\frac14 E^{(4)} R^{(4)}
+\frac{1}{32}  E^{(4)} \partial_{\bar{\mu}}\ln\Delta\; 
\partial^{\bar{\mu}}\ln\Delta
-\frac1{16}  E^{(4)} \partial_{\bar{\mu}}g_{\bar{m}\bar{n}}\; 
\partial^{\bar{\mu}}g^{\bar{m}\bar{n}}
\n
&&
-\frac1{12}  E^{(4)} \partial_{\bar{\mu}}
A_{(\bar{i}+2)(\bar{j}+2)(\bar{k}+2)}\; 
\partial^{\bar{\mu}}
A^{(\bar{i}+2)(\bar{j}+2)(\bar{k}+2)}
\n
&&
+\frac1{16} E^{(4)} \; \sqrt{\Delta}
B^{(4) \bar{i}}_{\bar{\mu}\bar{\nu}} 
B_{\bar{i}}^{(4) \bar{\mu}\bar{\nu}}
-\frac1{12}  E^{(4)}  \Delta\; 
F^{(4)}_{\bar{\mu}\bar{\nu}\bar{\rho}\; \bar{i}} 
F^{(4) \bar{\mu}\bar{\nu}\bar{\rho}\; \bar{i}}
-\frac1{8}  E^{(4)}  \sqrt{\Delta}\;
F^{(4)}_{\bar{\mu}\bar{\nu}\; \bar{i}\bar{j}} 
F^{(4) \bar{\mu}\bar{\nu}\; \bar{i}\bar{j}} 
\n
&&
-\frac2{12^3}
\epsilon^{\bar{\mu}\bar{\nu}\bar{\rho}\bar{\sigma}}
\epsilon^{\bar{i}\bar{j}\bar{k}\bar{l}\bar{m}\bar{n}\bar{o}}
\Big(\ 4 F^{(4)}_{\bar{\mu}\bar{\nu}\bar{\rho}\; \bar{i}} 
\partial_{\bar{\rho}} 
A_{(\bar{j}+2)(\bar{k}+2)(\bar{l}+2)}
A_{(\bar{m}+2)(\bar{n}+2)(\bar{o}+2)}
\n
&&
\hspace*{3.2cm} 
-9 F^{(4)}_{\bar{\mu}\bar{\nu}\; \bar{i}\bar{j}}
F^{(4)}_{\bar{\rho}\bar{\sigma}\; \bar{k}\bar{l}}
A_{(\bar{m}+2)(\bar{n}+2)(\bar{o}+2)}
\n
&&
\hspace*{3.2cm}
+
9 F^{(4)}_{\bar{\mu}\bar{\nu}\; \bar{i}\bar{j}}
B^{(4) \bar{p}}_{\bar{\rho}\bar{\sigma}}
A_{(\bar{p}+2)(\bar{k}+2)(\bar{l}+2)}
A_{(\bar{m}+2)(\bar{n}+2)(\bar{o}+2)}
\n
&&
\hspace*{3.2cm}
-
3 B^{(4) \bar{p}}_{\bar{\mu}\bar{\nu}}
B^{(4) \bar{q}}_{\bar{\rho}\bar{\sigma}}
A_{(\bar{p}+2)(\bar{i}+2)(\bar{j}+2)}
A_{(\bar{q}+2)(\bar{k}+2)(\bar{l}+2)}
A_{(\bar{m}+2)(\bar{n}+2)(\bar{o}+2)}
\Big)\n
\label{4de7}
\eeqa

\noi
where the elfbein is parameterized as

\beq
E_{~M}^{(11)A}=\left[
\begin{array}{cc}\Delta^{-\frac14}
E_{~\bar{\mu}}^{(4)\bar{\alpha}}
& B_{\bar{\mu}}^{(4) \bar{i}} \rho^{\ \bar{a}}_{\bar{i}}\\
0&\rho_{\bar{i}}^{\ \bar{a}}\end{array}
\right]
\label{4de7metr}
\eeq

\noi
with the internal triangular 
vielbein $\rho_{\bar{i}}^{\ \bar{a}}$. 
$\bar{i},\ \bar{a}$ are curved and flat internal indices 
respectively. They have been chosen to run in $2...8$ 
for later convenience. The internal metric 
$g_{\bar{m}\bar{n}}$ is defined as usual to be 
$g_{\bar{m}\bar{n}}=
\rho_{\bar{m}\; \bar{a}}\rho_{\bar{n}}^{\ \bar{a}}$ and has 
signature $--\dots -$. Its determinant is
$\sqrt{\Delta}=\det \rho_{\bar{m}}^{\ \bar{a}}$.

$\bar{\mu},\bar{\nu},\ldots$ are curved and $\bar{\alpha},
\bar{\beta},\ldots$
flat four dimensional indices. They run from $0\dots 3$.
For the fields $F^{(4)}_{\bar{\mu}\bar{\nu}\; \bar{i}\bar{j}}$
and $F^{(4)}_{\bar{\mu}\bar{\nu}\bar{\rho}\; \bar{i}}$, the 
following definitions were used in order to ensure the suitable
transformation properties with respect to 
internal diffeomorphisms:

\beqa
F^{(4)}_{\bar{\mu}\bar{\nu}\; \bar{i}\bar{j}} &=& 
F'_{\bar{\mu}\bar{\nu}\; \bar{i}\bar{j}} 
+ B^{(4) \bar{k}}_{\bar{\mu}\bar{\nu}} 
A_{(\bar{i}+2)(\bar{j}+2)(\bar{k}+2)}
\n
F'_{\bar{\mu}\bar{\nu}\; \bar{i}\bar{j}}
&=&2\partial_{[\bar{\mu}}A'_{\bar{\nu}]\; 
(\bar{i}+2)(\bar{j}+2)}, 
\n
A'_{\bar{\mu}\; (\bar{i}+2)(\bar{j}+2)}
&=&
A_{\bar{\mu}\; (\bar{i}+2)(\bar{j}+2)}
-B^{(4) \bar{k}}_{\bar{\mu}}
A_{(\bar{i}+2)(\bar{j}+2)(\bar{k}+2)},
\n
B^{(4) \bar{k}}_{\bar{\mu}\bar{\nu}} 
&=& 2 \partial_{[{\bar{\mu}}} 
B^{(4) \bar{k}}_{\bar{\nu}]},
\n
F^{(4)}_{\bar{\mu}\bar{\nu}\bar{\rho}\; \bar{i}} &=&
F'_{\bar{\mu}\bar{\nu}\bar{\rho}\; \bar{i}} 
+ 3 B^{(4) \bar{k}}_{[\bar{\mu}\bar{\nu}} 
A_{\bar{\rho}](\bar{i}+2)(\bar{k}+2)}
\n
F'_{\bar{\mu}\bar{\nu}\bar{\rho}\; \bar{i}}
&=& 3\partial_{[\bar{\mu}} 
A'_{\bar{\nu}\bar{\rho}]\; (\bar{i}+2)},
\n 
A'_{\bar{\mu}\bar{\nu}\; (\bar{i}+2)}&=&
A_{\bar{\mu}\bar{\nu}\; (\bar{i}+2)}
- 2 B^{(4) \bar{j}}_{[\bar{\mu}}
A_{\bar{\nu}](\bar{i}+2)(\bar{j}+2)}
-B^{(4) \bar{j}}_{\bar{\mu}}B^{(4) \bar{k}}_{\bar{\nu}}
A_{(\bar{j}+2)(\bar{k}+2)(\bar{i}+2)}.
\nonumber
\eeqa

\noi 
The fields $F'_{\bar{\mu}\bar{\nu}\; \bar{i}\bar{j}}$ and 
$F'_{\bar{\mu}\bar{\nu}\bar{\rho}\; \bar{i}}$
are then dualized by adding

\bdm
{\cal L}_{\rm Lag.mult.}=\frac1{12}
\varphi^{(4) \bar{i}}
\epsilon^{\bar{\mu}\bar{\nu}\bar{\rho}\bar{\sigma}}
\partial_{\bar{\mu}}
F'_{\bar{\nu}\bar{\rho}\bar{\sigma}\; \bar{i}}
+\frac14 \tilde{A}_{\bar{\sigma}}^{\bar{i}\bar{j}}
\epsilon^{\bar{\mu}\bar{\nu}\bar{\rho}\bar{\sigma}}
\partial_{\bar{\rho}}
F'_{\bar{\mu}\bar{\nu}\; \bar{i}\bar{j}}.
\edm

\noi
In order to simplify the Lagrangian and make 
the hidden $E_{7(+7)}$ symmetry manifest, 
the new field strength 
${\tilde A}^{\bar{i}\bar{j}}_{\bar{\mu}\bar{\nu}}
= 2 \partial_{[{\bar{\mu}}}
{\tilde A}^{\bar{i}\bar{j}}_{\bar{\nu}]}$
is introduced, as well as the vector notation

\bdm
{\cal G}_{\bar{\mu}\bar{\nu}}
=
\left[\begin{array}{c}
\tilde{A}^{\bar{i}\bar{j}}_{\bar{\mu}\bar{\nu}}\\
\tilde{A}^{\bar{i} 9}_{\bar{\mu}\bar{\nu}}\end{array} \right],\
{\cal H}_{\bar{\mu}\bar{\nu}}=
\left[\begin{array}{c}
H^{\tilde{A}}_{\bar{\mu}\bar{\nu}\; \bar{i}\bar{j}}\\
H^{\tilde{A}}_{\bar{\mu}\bar{\nu}\; \bar{i}9}\end{array} \right]
\edm

\noi
where $\tilde{A}^{\bar{i} 9}_{\bar{\mu}\bar{\nu}}=
-{\scriptstyle\frac12}B^{(4) \bar{i}}_{\bar{\mu}\bar{\nu}}$.
The dual fields ${\cal H}_{\bar{\mu}\bar{\nu}}$ obey

\bdm
H^{\tilde{A}}_{\bar{\mu}\bar{\nu}\;\bar{i}\bar{j}}=
-\frac4{E^{(4)}}\; \star \left(\frac{\delta{\cal L}}{\delta
{\tilde A}^{\bar{\mu}\bar{\nu}\;\bar{i}\bar{j}}}\right),\ \
H^{\tilde{A}}_{\bar{\mu}\bar{\nu}\;\bar{i} 9}=
-\frac4{E^{(4)}}\; \star \left(\frac{\delta{\cal L}}{\delta
\tilde{A}^{\bar{\mu}\bar{\nu}\;\bar{i} 9}}\right).
\edm

\noi
The 70 scalars of the theory, 
$A_{(\bar{i}+2)(\bar{j}+2)(\bar{k}+2)}$,
$\rho_{\bar{i}}^{\ \bar{a}}$ and
$\varphi^{(4)\; \bar{i}}$, 
are joint together in a field 
${\cal V}^{(4)}\in E_{7(+7)}/SU(8)$, a representation matrix 
in the fundamental {\bf 56} 
representation \cite{Cre79}. 

The fundamental {\bf 56} representation is 
given in the following way: The representation 
space is spanned by two antisymmetric 
tensors $x^{\hat{i}\hat{j}}, y_{\hat{i}\hat{j}}$, where indices
$\hat{i},\hat{j}$ are chosen to run in $2\dots 9$. 
On a vector 
$(x^{\hat{i}\hat{j}} | y_{\hat{i}\hat{j}})^t$, 
the algebra 
$\frak{e}_{7(+7)}$ acts by the real matrices

\beqa
\Lambda&=&\left[
\begin{array}{cc}
2\Lambda^{[\hat{i}}_{~~[\hat{k}}\delta^{\hat{j}]}_{\hat{l}]}&\\
&2\Lambda_{[\hat{i}}^{~~[\hat{k}}\delta_{\hat{j}]}^{\hat{l}]}
\end{array}
\right],\n
\Sigma&=&\left[
\begin{array}{cc}
&\Sigma^{*\hat{i}\hat{j}\hat{k}\hat{l}}\\
\Sigma_{\hat{i}\hat{j}\hat{k}\hat{l}}&
\end{array}
\right]
\label{cartan}
\eeqa 

\noi
where 
$\Lambda^{\hat{i}}_{\ \hat{k}}= -\Lambda^{\ \hat{i}}_{\hat{k}}$,
$\Lambda^{\hat{i}}_{\ \hat{i}}=0$ obviously represents a
$\frak{sl}(8)$ subalgebra, and 
$\Sigma_{\hat{i}\hat{j}\hat{k}\hat{l}}$ is totally antisymmetric.
We have
$\Sigma^{*\hat{i}\hat{j}\hat{k}\hat{l}}=
{\scriptstyle\frac{1}{24}}
\epsilon^{\hat{i}\hat{j}\hat{k}\hat{l}\hat{m}\hat{n}
\hat{o}\hat{p}}
\Sigma_{\hat{m}\hat{n}\hat{o}\hat{p}}$. 

The representation (\ref{cartan}) is symplectic, it preserves
the symplectic from

\bdm
\Omega=\left[\begin{array}{cc}
& -{\bf 1}^{\hat{i}\hat{j}\hat{k}\hat{l}} \\
{\bf 1}_{\hat{i}\hat{j}\hat{k}\hat{l}}&
\end{array}\right].
\edm 

\noi
We will now recover this representation of $\frak{e}_{7(+7)}$
embedded into the realization of 
$\frak{e}_{8(+8)}$ given in appendix A, using Freudenthal's
realization of exceptional Lie algebras \cite{Freudenthal}.

\begin{figure}[htbp]
\begin{center}
\leavevmode
\begin{picture}(0,0)%
\epsfig{file=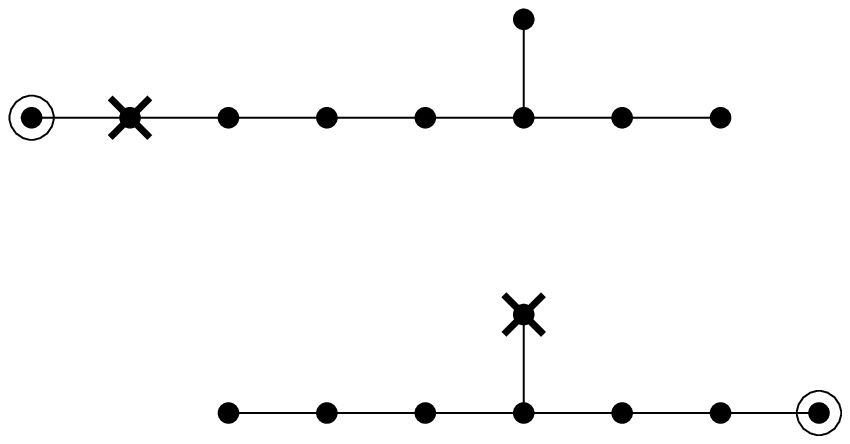}%
\end{picture}%
\setlength{\unitlength}{4144sp}%
\begin{picture}(3882,2337)(901,-2617)
\put(2431,-2581){\makebox(0,0)[lb]{\footnotesize$\eps_{34}$}}
\put(2881,-2581){\makebox(0,0)[lb]{\footnotesize$\eps_{45}$}}
\put(3331,-2581){\makebox(0,0)[lb]{\footnotesize$\eps_{56}$}}
\put(3781,-2581){\makebox(0,0)[lb]{\footnotesize$\eps_{67}$}}
\put(4231,-2581){\makebox(0,0)[lb]{\footnotesize$\eps_{78}$}}
\put(1981,-2581){\makebox(0,0)[lb]{\footnotesize$\eps_{23}$}}
\put(3286,-1726){\makebox(0,0)[lb]{\footnotesize$\eps_{678}$}}
\put(2431,-1231){\makebox(0,0)[lb]{\footnotesize$\eps_{34}$}}
\put(2881,-1231){\makebox(0,0)[lb]{\footnotesize$\eps_{45}$}}
\put(3331,-1231){\makebox(0,0)[lb]{\footnotesize$\eps_{56}$}}
\put(3781,-1231){\makebox(0,0)[lb]{\footnotesize$\eps_{67}$}}
\put(4231,-1231){\makebox(0,0)[lb]{\footnotesize$\eps_{78}$}}
\put(4681,-2581){\makebox(0,0)[lb]{\footnotesize$\eps_{189}$}}
\put(3286,-376){\makebox(0,0)[lb]{\footnotesize$\eps_{678}$}}
\put(901,-1231){\makebox(0,0)[lb]{\footnotesize$-\eps_{19}$}}
\put(1486,-1231){\makebox(0,0)[lb]{\footnotesize$\eps_{12}$}}
\put(1981,-1231){\makebox(0,0)[lb]{\footnotesize$\eps_{23}$}}
\end{picture}
\end{center}
\caption{\label{e8decomp} Decomposition of 
$E_{8(+8)}\supset SL(2)\times E_{7(+7)}$ and
$E_{7(+7)}\supset SL(8)$}
\end{figure}

From the decomposition illustrated in figure 
\ref{e8decomp} we learn that the $\frak{e}_{7(+7)}$ subalgebra 
of $\frak{e}_{8(+8)}$ may be generated by  

\beq
h_{1\bar{i}9},\
E^{1\bar{i}9},\
E^*_{1\bar{i}9},\
E^{\bar{i}}_{~\bar{j}},\
E^{\bar{i}\bar{j}\bar{k}},\
E^*_{\bar{i}\bar{j}\bar{k}}.
\label{e7generators}
\eeq

\noi
Let us define 
\beqa
\Lambda_{\frak{e}_{8(+8)}} &=& \sum_{\bar{i}=2}^8
\left(
\Lambda_{\bar{i}}^{~\bar{i}}h_{1\bar{i}9}
+\Lambda_{\bar{i}}^{~9}E^{1\bar{i}9}
+\Lambda_{9}^{~\bar{i}}E^*_{1\bar{i}9}
\right)
+ \sum_{\bar{i},\bar{j}=2}^8
\Lambda_{\bar{i}}^{~\bar{j}}
E^{\bar{i}}_{~\bar{j}},\n
\Sigma_{\frak{e}_{8(+8)}}&=&
\sum_{\bar{i},\bar{j},\bar{k}=2}^8\frac2{3!}\left(
\Sigma_{\bar{i}\bar{j}\bar{k}9}
E^{\bar{i}\bar{j}\bar{k}}
+\Sigma^{*\bar{i}\bar{j}\bar{k}9}
E^*_{\bar{i}\bar{j}\bar{k}}
\right).
\label{rep1}
\eeqa

\noi 
For the representation space basis, we define the vectors

\beqa
{\cal S}^t &=&\Big(-E^*_{\bar{i}\bar{j}9},
+E^1_{~\bar{i}}\; | \;
-E^{1\bar{i}\bar{j}}, - E^{\bar{i}}_{~9}\Big),
\n
{\cal X}^t &=&\Big( x^{\bar{i}\bar{j}}, x^{\bar{i}\bar{9}} 
\; | \;
 y_{\bar{i}\bar{j}}, y_{\bar{i}\bar{9}} \Big).
\label{rep2}
\eeqa

\noi
Using the relations among the generators 
(\ref{E8generatorrelations}), 
one may verify that 

\beq
[\Lambda_{\frak{e}_{8(+8)}}, 
{\cal X}\cdot{\cal S}]
= {\cal X}'\cdot{\cal S},\ \ \ \ 
[\Sigma_{\frak{e}_{8(+8)}},
{\cal X}\cdot{\cal S}]
= {\cal X}''\cdot{\cal S}
\label{rep3}
\eeq 

\noi
with

\bdm
{\cal X}'= \Lambda \cdot {\cal X},\ \ \ \ 
{\cal X}''= \Sigma \cdot {\cal X} 
\edm

\noi
reproducing exactly the action of (\ref{cartan}). 
We can therefore use the 
adjoint action on ${\cal S}$ to define the {\bf 56} 
representation ${\bf \rho_{56}}$ of 
$\frak{e}_{7(+7)}$ as subalgebra of 
$\frak{e}_{8(+8)}$.

Later we use $i,j,\ldots=1,2,\ldots,8$ as 
the indices for the eight dimensional torus in the 
dimensional reduction to $d=3$. Note that the $SL(8,\RR)$ 
subgroup  
generated by $\Lambda$'s is {\it not} the modular group of this 
torus.

The scalar field ${\cal V}^{(4)}$ is given explicitly by

\beqa
{\cal V}^{(4)} &=&
\ \exp\left(-\sum_{\bar{m}=2}^{8}
\ln \left(\rho_{\bar{m}}^{\ \dot{\bar{m}}}\right)\ h_{1\bar{m}9}
+ \frac18 \ln \Delta h_{1\bar{m}9} 
\right)
\n
&&
\ \ \ \ \ \ \prod_{\bar{p}=0}^{5}
\exp \left(-\sum_{\bar{q},
\bar{r}=8-\bar{p}}^8 \rho_{7-\bar{p}}^{\ \ \ \ \ \dot{\bar{q}}}
(\rho^{(7-\bar{p})\ -1})^{\ \ \bar{r}}_{\dot{\bar{q}}}
\ E^{7-\bar{p}}_{\ \ \ \  \ \bar{r}} \right)
\n            
&&
\ \ \ \ \ \ \exp\left(-\sum_{\bar{i}=2}^{8}\varphi^{(4)\
\bar{i}}\
E^*_{1\bar{i}9} \right)
\n
&&
\ \ \ \ \ \ \exp
\left(\frac{2}{3!} \sum_{\bar{i},\bar{j},\bar{k}=2}^8
A_{(2+\bar{i})(2+\bar{j})(2+\bar{k})}\ E^{\bar{i}\bar{j}\bar{k}}
\right)
\label{ve7}
\eeqa       

\noi
in the above ${\bf \rho_{56}}$ representation, 
where $\rho_{\bar{i}}^{(n)\; \bar{a}}$ is the submatrix 
of $\rho_{\bar{i}}^{\ \bar{a}}$ with columns and rows 
$(n+1)\dots 8$. 
This result can be seen to correspond 
to the one in \cite{Cre98}.

From ${\cal V}^{(4)}$, again 
a field $P^{(4)}_\mu$ may be defined by

\bdm
\partial_{\mu}{\cal V}^{(4)} {\cal V}^{(4) -1}= 
Q^{(4)}_{\mu} + P^{(4)}_{\mu},
\ \ Q^{(4)}_{\mu} \in \frak{su}(8),
\ \ P^{(4)}_{\mu} \in \frak{e}_{7(+7)}-\frak{su}(8).
\edm

\noi
The vector fields are related by the twisted self-duality 
relation
\bdm
{\cal F}_{\bar{\mu}\bar{\nu}} \equiv
\left[\begin{array}{c}
{\cal G}_{\bar{\mu}\bar{\nu}} \\ {\cal H}_{\bar{\mu}\bar{\nu}}
\end{array}\right]
=\Omega{\cal V}^{(4)T}{\cal V}^{(4)}
\left[\begin{array}{c}
\star {\cal G}_{\bar{\mu}\bar{\nu}} \\
\star {\cal H}_{\bar{\mu}\bar{\nu}}
\end{array}\right].
\edm

\noi
The Lagrangian finally takes the form

\beq
{\cal L}
=  - \frac14 E^{(4)} R^{(4)}+ 
E^{(4)} \left(\frac18 {\cal G}^T_{\bar{\mu}\bar{\nu}}
\star {\cal H}^{\bar{\mu}\bar{\nu}}
+ \frac1{48}{\rm Tr}(P^{(4)}_{\bar{\mu}}
P^{(4) \bar{\mu}})\right).
\label{4dact}
\eeq

The scalar part of the action is invariant under
${\cal V}^{(4)}\rightarrow h(x){\cal V}^{(4)}\Lambda$, 
$\Lambda\in E_{7(+7)}$,
$h(x)\in SU(8)$, where the local $SU(8)$
is used to restore the parameterization of coset space. 

The combined vector field ${\cal F}$ 
transforms as a vector with respect 
to $E_{7(+7)}$, but not the entire $E_{7(+7)}$ is a 
symmetry of the action.
Writing the
vector field part of the action as

\bdm
\frac1{16} E^{(4)} \Big({\cal F}^T_{\bar{\mu}\bar{\nu}}
\;\;L\;
\star {\cal F}^{\bar{\mu}\bar{\nu}} 
\Big)
\edm

\noi
with

\bdm
L =\left[\begin{array}{cc}
& {\bf 1}^{\hat{i}\hat{j}\hat{k}\hat{l}} \\
{\bf 1}_{\hat{i}\hat{j}\hat{k}\hat{l}}&
\end{array}\right]
\edm

\noi
we see that a symmetry of the Lagrangian has to
preserve $L$. 
It follows that only $\Lambda^{\hat{i}}_{\ \hat{k}}$ 
in (\ref{cartan}) can be nonzero, the action is invariant under 
the $SL(8)$ subgroup in figure \ref{e8decomp}. We will treat 
the symmetries and subgroups more closely in the following 
section. 

\subsection{Duality in $d=4$}

The equations of motion of the theory show the full
$E_{7(+7)}$ invariance.
The classical duality 
symmetry is given by
  
\bdm
{\cal F}_{\hat{\mu}\hat{\nu}}\to\Lambda^{-1}
{\cal F}_{\hat{\mu}\hat{\nu}},\ \
{\cal V}^{(4)}\rightarrow {\cal V}^{(4)} \Lambda,
\ \ \Lambda \in E_{7(+7)}.
\edm

\noi
Again, the 
transformation of ${\cal V}^{(4)}$ is accompanied by a 
compensating
local $SU(8)$ transformation
${\cal V}^{(4)}\rightarrow h(x){\cal V}^{(4)}\Lambda$,
$h(x)\in SU(8)$ to restore the parameterization
of the coset space.

We again define charges 

\bdm
{\cal Z}= \left[\begin{array}{c}
p \\
q
\end{array}\right],\ \ \
p= \frac{1}{2\pi}\oint_\Sigma {\cal G}, \ \ \
q= \oint_\Sigma {\cal H}.
\edm

\noi
It has been argued in \cite{Hul95} that all magnetic and 
electric charges exist. It is actually clear from the basis 
(\ref{rep2}) and the fact that the {\bf 56} representation of 
$E_{7(+7)}$ is minimal that, if a solution with one 
nonzero charge exists, solutions with a single 
charge carried by all other gauge fields may be obtained 
by Weyl reflections that are group elements of $E_{7(+7)}$, to be 
introduced below.

The DSZ condition 

\bdm
{\cal Z}^t \;\Omega\; {\cal Z}'= n,\ n\in \ZZ
\edm

\noi
breaks $E_{7(+7)}$ to $E_{7(+7)}(\ZZ)$, demanding integer 
shifts on the lattice defined by the basis (\ref{rep2}). This 
group has been proposed to be a unified duality symmetry of
type II string theory in \cite{Hul95}, called U-duality for 
short, unifying strong-weak coupling dualities and 
target space dualities and putting all 70 moduli of the 
theory, including the string coupling constant,
on the same footing. We will analyse the subgroups corresponding 
to T- and S-duality in our notation after introducing generators 
for the discrete group.

What are generators of this $E_{7(+7)}(\ZZ)$? 
It may be checked that the
basis ${\cal S}$ in (\ref{rep2}) forms an admissible lattice 
(see appendix B) of 
$\frak{e}_{7(+7)}$ in the representation ${\bf \rho_{56}}$ 
defined by (\ref{rep1}),(\ref{rep2}),(\ref{rep3}). 
The {\bf 56} representation of 
$\frak{e}_{7(+7)}$ is furthermore the unique minimal 
representation \cite{Humphreys}. From the proof in appendix B 
it follows that the subgroup inducing integer shifts on the
lattice with base ${\cal S}$, called $E_{7(+7)}(\ZZ)$ in the 
following, is generated by ``fundamental unipotents'', that is,

\beqa
T^{\bar{i}}_{\ \bar{j}}=
\exp(E^{\bar{i}}_{\ \bar{j}}),
\ \ \bar{i}<\bar{j}\ &;& \  
T^{\bar{i}}_{\ \bar{j}}=
\exp(E^{\bar{i}}_{\ \bar{j}}),
\ \ \bar{i}>\bar{j},
\n
T^{1\bar{i}9}=
\exp(E^{1\bar{i}9})
\ &;& \  
T^*_{1\bar{i}9}=
\exp(E^*_{1\bar{i}9}),
\n
T^{\bar{i}\bar{j}\bar{k}}=
\exp(E^{\bar{i}\bar{j}\bar{k}})
\ &;& \  
T^*_{\bar{i}\bar{j}\bar{k}}=
\exp(E^*_{\bar{i}\bar{j}\bar{k}})
\label{discrete1}
\eeqa

\noi 
or alternatively 

\beqa
T^{\bar{i}}_{\ \bar{j}}=
\exp(E^{\bar{i}}_{\ \bar{j}}),
\ &;& \  
S^{\bar{i}}_{\ \bar{j}}=
\exp(-E^{\bar{j}}_{\ \bar{i}})
\exp(E^{\bar{i}}_{\ \bar{j}})\exp(-E^{\bar{j}}_{\ \bar{i}})
\ \ \bar{i}<\bar{j},
\n
T^{1\bar{i}9}=
\exp(E^{1\bar{i}9})
\ &;& \  
S^{1\bar{i}9}=
\exp(-E^*_{1\bar{i}9})\exp(E^{1\bar{i}9})
\exp(-E^*_{1\bar{i}9}),
\n
T^{\bar{i}\bar{j}\bar{k}}=
\exp(E^{\bar{i}\bar{j}\bar{k}})
\ &;& \  
S^{\bar{i}\bar{j}\bar{k}}=
\exp(-E^*_{\bar{i}\bar{j}\bar{k}})
\exp(E^{\bar{i}\bar{j}\bar{k}})
\exp(-E^*_{\bar{i}\bar{j}\bar{k}})
\label{discrete2}
\eeqa 

\noi
in the representation ${\bf \rho_{56}}$,
where the S generators are known to carry a representation 
of the Weyl group modulo $\ZZ_2$ (see e.g. \cite{Ler89}).

This yields a complete set of discrete generators of the 
$E_{7(+7)}(\ZZ)$ U-duality. U-duality groups in higher 
dimensions are found by their embeddings.
Since the notion of the above generators is representation 
independent, the $E_{6(+6)}(\ZZ)$ etc. U-duality generators 
follow directly from truncating the Dynkin diagram.
Furthermore, the corresponding representations in these 
truncations are always minimal, and 
the corresponding discrete group generator representations
can be read off explicitly from the basis (\ref{rep2}). 
Actually, in this sense, all U-dualities follow from the adjoint
representation of $E_{8(+8)}$.

The equations 
(\ref{rep1}), (\ref{rep2}), (\ref{rep3}) 
together with (\ref{ve7}) and the definition of ${\cal F}$
enable us to give
U-duality transformations explicitly in $d=4$, but give a
direct contact to the algebraic notations of \cite{Eli97} as well.
We will illustrate this below.

It is instructive to identify 
T- and S-duality symmetries in 
the theory along the lines of \cite{Sen94, Sen95}.

The known superstring theories in ten dimensions have a common 
low-energy sector whose spectrum is the same 
as the NS-NS sector of type II theories. 
The corresponding
low-energy fields will be called 
NS-NS fields. 

In $d=4$, the fundamental 
string can carry electric charge with respect to the U(1) fields
in the NS-NS sector \cite{Nar86}.
T-duality is identified with the subgroup of $E_{7(+7)}$ 
that stabilizes the NS-NS charge lattice.

The NS-NS sector in $d=10$ consists of the metric, 
the dilaton and an antisymmetric two-form. 
Looking at (\ref{11d}) and considering the direction 
10 as eleventh spatial dimension, the NS-NS fields
may be identified with

\beqa
&E_{\bar{\mu}}^{(4)\bar{\alpha}},\
\rho_8^{\ \dot{8}},\ 
\rho_{\tilde{i}}^{\ \tilde{a}},\ 
B^{(4)\tilde{i}}_{\mu},\n
& 
A_{\bar{\mu}\bar{\nu}8},\ 
A_{\bar{\mu}\tilde{i}8},\ 
A_{\tilde{i}\tilde{j}8}
\nonumber
\eeqa

\noi
where indices with tilde run from $2\dots 7$.

The $d=4$ U(1) field strengths in the NS-NS sector are

\bdm
B^{(4)\ \tilde{i}}_{\bar{\mu}\bar{\nu}}= 
-2 \tilde{A}_{\bar{\mu}\bar{\nu}}^{\tilde{i}9},\ \ 
F'_{\bar{\mu}\bar{\nu}\ \tilde{i}8}= -H_{\bar{\mu}\bar{\nu}
\ \tilde{i}8},\ \
\tilde{i}\in \{2,\dots 7\}.
\edm

\noi
where we assume that the RR fields are set to zero.

With (\ref{rep2}) this corresponds to the 
representation space basis

\beq
(E^{1\tilde{i}8},\ \ -E^1_{\ \tilde{i}}).
\label{o6}
\eeq

\begin{figure}[htbp]
\begin{center}
\leavevmode
\begin{picture}(0,0)%
\epsfig{file=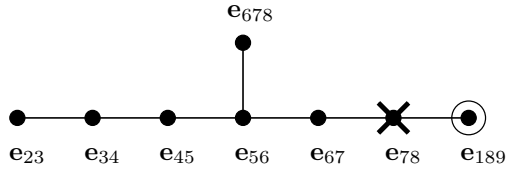}%
\end{picture}%
\setlength{\unitlength}{4144sp}%
\begin{picture}(2862,987)(1973,-2617)
\put(2431,-2581){\makebox(0,0)[lb]{\footnotesize$\eps_{34}$}}
\put(2881,-2581){\makebox(0,0)[lb]{\footnotesize$\eps_{45}$}}
\put(3331,-2581){\makebox(0,0)[lb]{\footnotesize$\eps_{56}$}}
\put(3781,-2581){\makebox(0,0)[lb]{\footnotesize$\eps_{67}$}}
\put(4231,-2581){\makebox(0,0)[lb]{\footnotesize$\eps_{78}$}}
\put(1981,-2581){\makebox(0,0)[lb]{\footnotesize$\eps_{23}$}}
\put(3286,-1726){\makebox(0,0)[lb]{\footnotesize$\eps_{678}$}}
\put(4681,-2581){\makebox(0,0)[lb]{\footnotesize$\eps_{189}$}}
\end{picture}
\end{center}
\caption{\label{e7decomp} Decomposition of 
$E_{7(+7)}\supset O(6,6)\times SL(2)$}
\end{figure}

\noi
Using (\ref{E8generatorrelations}) one may verify that the 
subgroup 
stabilizing this basis is the obvious $O(6,6)$ subgroup  
generated as indicated in figure 
\ref{e7decomp}. 

\noi
With the above representation 
basis one may verify that it acts in the fundamental 
representation and  
preserves the metric

\beq
\left[\begin{array}{cc}
& {\bf 1} \\
{\bf 1}&
\end{array}\right]
\label{presmetr}
\eeq

\noi 
where {\bf 1} is the six dimensional unit matrix. 
The dual fields

\bdm
-\frac12 H_{\bar{\mu}\bar{\nu}\ \tilde{i}9},\ \ 
-\tilde{A}_{\bar{\mu}\bar{\nu}}^{\tilde{i}8},\ \
\tilde{i}\in \{2,\dots 7\}.
\edm  

\noi
correspond to the basis

\bdm
(E^*_{\tilde{i}89},\ \ E^{\tilde{i}}_{\ 9})
\edm

\noi
and transform in the transpose inverse representation of the 
above fundamental representation of $O(6,6)$ as expected.
We explicitly have the decomposition

\beqa
E_{7(+7)}&\supset& O(6,6)\times SL(2,\RR)
\n
{\bf 56}&\supset&({\bf 12},{\bf 2}) + ({\bf 32},{\bf 1})
\nonumber
\eeqa

\noi 
mentioned in \cite{Hul95}.
The $U(1)$ field strength in the R-R sector transform in 
the ${\bf 32}$ spinor representation of $O(6,6)$, which means 
that T-duality mixes electric and magnetic fields in the R-R 
sector.

\noi 
The action of $O(6,6)$ on the Cartan subalgebra part of (\ref{ve7})
can be seen by looking at the submatrix of (\ref{ve7}) 
corresponding to the basis (\ref{o6}).
This yields a matrix 
with the $\rho_m^{\ \dot{m}} \Delta^{-\frac18}$ on the diagonal.
The $\rho_m^{\ \dot{m}}$ correspond to the radii of the 
compactification torus transforming under $O(6,6)$, 
while $\Delta^{1/2}$ is the volume of 
the 7-torus including the eleventh direction. The 
appearance of the factor $\Delta^{-1/8}$
indicates that $\rho_8^{\ \dot{8}}$, corresponding to the radius of
the eleventh compact direction, 
transforms as well under T-duality and 
is not decoupled, as was pointed out e.g. in \cite{Obe98}. This
transformation therefore mixes strong weak coupling duality 
of the type IIB string \cite{Sch95, Asp96}, that corresponds to 
modular transformations
involving $\rho_8^{\ \dot{8}}$,  with a transformation of
the moduli corresponding to the compactification torus of type II
string theory.

Note that the $O(6,6)$ in our formulation is not a symmetry of 
the action, but only of the equations of motion.  

We now turn to the commuting 
$SL(2,\RR)$ factor in figure \ref{e7decomp}, which is a 
symmetry of the action (\ref{4dact})\footnote{This is 
analogous to 
Sen's manifestly $SL(2,\RR)$ invariant
action from the dual $N=1$ $d=10$ supergravity theory. 
The connection is obvious when all R-R fields are set to zero.}. 
It was suggested
in \cite{Hul95} to interpret this factor like in the
heterotic case \cite{Sen94} as a $d=4$
S-duality, not to be confused with the $d=10$ S-duality of the 
type IIB string.  

Interpreting the above $SL(2,\RR)$ factor as S-duality, 
a $\ZZ_2$ symmetry exchanging electric 
and magnetic sector is expected to be present. 
The natural candidate is $S^{189}$. Using 
(\ref{E8generatorrelations}), it is interesting to note 
that $S^{189}$ transforms
the magnetic into the electric sector and vice versa, but
takes the Kaluza-Klein sector into the 3-form field sector 
(and vice versa) as well.
In the basis we have 
chosen, $S^{189}$ has 
to be accompanied by an $O(6,6)$ transformation in order to 
transform the magnetic to the electric field strength of a 
specific vector field. 
This corresponds to the above
preserved metric (\ref{presmetr}), which is equal to

\bdm
\prod_{\tilde{i}<\tilde{j}} S^{\tilde{i}}_{\ \tilde{j}}\ 
\prod_{\tilde{i}<\tilde{j}} S^{\tilde{i}\tilde{j}8}.
\edm 
 
\noi
A suitable choice of basis can prevent us from
needing this additional $O(6,6)$ transformation. 

We will now compare our notion of 
$E_{7(+7)}(\ZZ)$ in (\ref{discrete1}), (\ref{discrete2}) 
with the definition
(\ref{orig2}) used in \cite{Obe98}.

In fact, since $E_7$ is simply laced, the 
generators in (\ref{discrete1}) corresponding to simple roots 
generate all other generators and therefore 
the whole $E_{7(+7)}(\ZZ)$. On the other hand, the same 
is true for the simply laced groups $O(6,6,\ZZ)$ and $SL(7,\ZZ)$ 
as subgroups of $E_{7(+7)}(\ZZ)$.
Both groups correspond to subdiagrams of the Dynkin diagram of 
$E_7$ and act in minimal representations. 
The subgroup $O(6,6)$ is indicated in figure \ref{e7decomp}, 
the subgroup $SL(7)$ simply corresponds to erasing the root
$\eps_{678}$ for the $E_7$ Dynkin diagram. 
Joining their generators together, we get the whole set of 
generators of $E_{7(+7)}(\ZZ)$.
The two definitions are therefore equivalent.

In the algebraic approach reviewed in
\cite{Obe98}, Weyl and Borel generators were used to 
define the discrete group. These actually correspond 
to the $S$ and $T$ generators in (\ref{discrete2}).
For the Weyl group, the identification

\bdm
S^{i}_{\ j}=\hat{S}_{ij},\ \ S^{ijk}=\hat{T}_{ijk}
\edm

\noi
holds, where hatted indices are the Weyl generators of \cite{Obe98}.
The $S^{i}_{j}$ correspond to the exchange of two radii, 
while the $S^{ijk}$ correspond to a simultaneous inversion of three
radii. Only the $S^{ij8}$ are elements of the T-duality group
$O(6,6)$, corresponding, as pointed out above, to a simultanous inversion of
two radii and the radius corresponding to the eleventh dimension
connected to type IIB S-duality. It should be kept in mind that the Weyl 
group representation carried by $S$ is a representation 
only modulo $\ZZ_2$ 
shifts on the Chevalley generators.  

The Borel generators are identified correspondingly, noting for the
T-duality group that
the $d=10$ antisymmetric tensor field corresponds to 
$A_{\tilde{i}\tilde{j}8}$. 

\subsection{Elementary Solitons}

In \cite{Hul95}, 
``elementary'' solitonic BPS solutions have been identified 
that each carry a single type of charge and therefore fill 
out an $E_{7(+7)}$ multiplet. This also accounts for the 
perturbative electrically charged string states that have 
their solitonic counterparts in these multiplets.
Note that these solitons in the NS-NS sector are exact classical solutions
to type II string theory \cite{Hul95}, not only of the low energy sector, and
therefore the whole multiplet consists of exact 
solutions. 

Other solutions 
corresponding to BPS solitons that break 1/4 and 1/8 
supersymmetry have been identified with multi-particle 
bound states of the ``elementary'' solitons charged with respect 
to different gauge fields \cite{Duf96}. Generically these 
correspond to intersecting p-brane configurations in
higher dimensions \cite{Pap96}. The most general 1/8 BPS soliton
is known to depend on five parameters, four charges and one 
relative phase (\cite{Ber99} and references therein).  

As already pointed out, the 
fact that the {\bf 56} representation is minimal guarantees that 
each gauge field can be reached
by a Weyl reflection 
from one nonzero gauge field.
We will sketch the well known $a=\sqrt{3}$ 
black hole solution \cite{Gib82} with $A_{MNP}=0$ in 
(\ref{11d}), which corresponds to the above 
solitons charged under one Kaluza-Klein gauge field, 
as an example. For the ten dimensional origin of this 
multiplet see e.g. \cite{Hul95} and \cite{Obe98}.

It is governed by one harmonic function $H,\ 
\partial_i\partial_i H =0$, where we have chosen Cartesian 
coordinates. The metric 
is of the form

\bdm
ds^2 = 
H^{-1/2} dt^2 - H^{1/2} dx_i^2
\edm

\noi
which directly resembles its origin as pp-wave of M-theory 
traveling in the compact direction that corresponds to the 
KK gauge field it is charged under.
Choosing 
a specific $\bar{k}\in\{1...7\}$, the electric 
solution reads in our notation

\bdm
B^{(4)\bar{k}}_{ti}
=
H^{-2} \partial_i H,\ \ \
g_{\bar{k}\bar{k}}= -H,
\edm

\noi
all other fields zero. The harmonic form is chosen to be

\bdm
H = 1 + \frac{Q}{|\vec{r}|}
\edm

\noi 
for a single BPS soliton, Q corresponds to the electric charge. 

For the corresponding multi black hole solution, the harmonic
form reads

\bdm
H = 1 + \sum_{n} \frac{Q_n}{|\vec{r}-\vec{r_n}|}
\edm

\noi 
reflecting the fact that the Bogomol'nyi bound is saturated 
and there are no forces between the constituents, allowing a 
superposition despite the nonlinearity of the Einstein 
equations.

The magnetic dual solution, corresponding to the wrapped 
M-theory KK6-brane, is given by

\bdm
B^{(4)\bar{k}}_{ij}
=
\epsilon_{ijk} \partial_k H,\ \ \
g_{\bar{k}\bar{k}}= -H^{-1},
\edm

\noi
with the harmonic form

\bdm
H = 1 + \frac{P}{|\vec{r}|}
\edm

\noi
for the single soliton with magnetic charge P, and

\bdm
H = 1 + \sum_{n} \frac{P_n}{|\vec{r}-\vec{r_n}|}
\edm

\noi
for the multi magnetically charged black hole.

Solutions charged with respect to the other gauge fields of 
the theory may now be obtained by the action
of $E_{7(+7)}$ resp. $E_{7(+7)}(\ZZ)$. They correspond 
to wrapped membrane and fivebrane solutions of M-theory. 
While the $S$ generators take single charge solutions to 
single charge solutions, the $T$
generators will yield solitons with multiple charge.

Note that the solitons may carry electric or magnetic charges, 
but there are no dyonic states in this multiplet.
Looking at (\ref{rep2}), one sees that no corresponding
$T$ generators exist. This is not surprising, since the 
dyonic solutions are known to have a different conformal 
structure, while U-duality does not act on the space-time 
metric.
 
We will now turn to the compactification to three dimensions.
We will assume as in \cite{Sen95} 
that the $E_{7(+7)}(\ZZ)$ 
symmetry is not broken. 

\subsection{The $d=3$ Theory}

The reduction to $d=3$ is strictly parallel to the toy model.
We will give the reduction here explicitly by using the 
embedding of $E_{7(+7)}$ in $E_{8(+8)}$ discussed above\footnote
{The $E_{8(+8)}$ symmetry in $d=3$ 
by reduction from $d=4$ was addressed in \cite{Jul82}.}.
 
Dropping the dependence on the compact coordinate,
we write the vierbein as 
\bdm
E_{~\bar{\mu}}^{(4)\bar{\alpha}}=\left[
\begin{array}{cc}e^{\phi/2} E_{~\mu}^{(3){\alpha}}
& e^{-\phi/2} \hat{B}_{\mu} \\
0&e^{-\phi/2}\end{array}
\right],
\edm     

\noi
where $\mu,\alpha$ now run from $0\dots 2$, 
and define the Kaluza-Klein invariant vector field strengths
${\cal G}'_{\mu\nu}=2\partial_{[\mu}{\cal G}'_{\nu]}$ 
with
${\cal G}'_{\mu}={\cal G}_{\mu}-\hat{B}_{\mu} {\cal G}_3$.
The vector fields are dualized by adding the Lagrange 
multipliers

\bdm
{\cal L}_{\rm Lag.mult.}=\frac14
\epsilon^{{\mu}{\nu}{\rho}}
\bar{\eta} \partial_{{\rho}}{\cal G}'_{{\mu}{\nu}}
+\frac{1}{8}
\epsilon^{{\mu}{\nu}{\rho}}
f \partial_{{\rho}}\hat{B}_{{\mu}{\nu}}.
\edm

\noi
With

\beqa
&\partial_{\mu} \eta = {\cal G}_{\mu 3},\ 
\partial_{\mu} \bar{\eta} = {\cal H}_{\mu 3},\ 
\cal{Y}=  \left[\begin{array}{c}
\eta \\
\bar{\eta}
\end{array}\right],&
\n
&\partial_{\mu} f = - \frac12 E^{(3)} 
\epsilon^{{\mu}{\nu}{\rho}} 
e^{-2\phi}
 \hat{B}_{{\nu}{\rho}} 
- {\cal Y}^t \Omega \partial_{\mu} {\cal Y}&
\nonumber
\eeqa

\noi
the Lagrangian gets

\beqa
{\cal L}
&=& -\frac14 E^{(3)} R^{(3)}
+\frac18  E^{(3)} \partial_{\mu}\phi\; \partial^{\mu}\phi
+\frac{1}{48}
E^{(3)}{\rm Tr}(P^{(4)}_{\mu}P^{(4) \mu})
\n
&&
+\frac18  E^{(3)} e^{2\phi} (\partial_{\mu} f + {\cal Y}^t 
\Omega \partial_{\mu}
 {\cal Y})
 (\partial^{\mu} f + {\cal Y}^t \Omega \partial^{\mu} {\cal Y})
\n
&&
+ \frac14 E^{(3)} e^{\phi} \partial_{\mu} {\cal Y}^t
{\cal V}^{(4) t} {\cal V}^{(4)} \partial^{\mu} {\cal Y}.
\label{e7}
\eeqa

\subsection{Elementary solitons in $d=3$}

We will now study how the $d=4$ solitons appear 
in the compactified theory.
For this, we study multi-BPS solitons where copies of a single
BPS soliton of the four dimensional theory are put
along the 3-direction with distance 
$2\pi R$ among two of them\footnote{Note 
that such solutions exist for the Schwarzschild 
case as well \cite{Kor95}.}.

This corresponds to the harmonic form

\bdm
H = 1 + \sum_{n} \frac{Z}
{\sqrt{x_1^2 + x_2^2 + (x_3 + 2 \pi R n)^2}}
\edm

\noi 
where $Z=Q,P$ for the electric resp. magnetic case. 

The above sum is logarithmically divergent, but the 
divergence can be regularized by adjusting an additive constant.
We can e.g. add a regulator of the form\footnote{We would like to
thank Hermann Nicolai for this comment.} 

\bdm
- \frac{Z}{2 \pi R n}.
\edm

\noi
Let us study the derivatives of H with respect to the 
cartesian coordinates.
It has been 
shown in \cite{Sen95} that the dependence of H on $x_3$ falls 
off exponentially if the compactification radius is small. 
To study the derivatives with respect to $x_1, x_2$, 
the summation is approximated by an integration 
\cite{Maldacena}, which finally yields

\bdm
H \propto \frac{Z}{2\pi R} \ln \rho + C
\edm 

\noi
where we introduced polar coordinates 
$\rho=\sqrt{x_1^2 + x_2^2}$ and 
$\theta$\footnote{See also \cite{Bla97}.}.
The result depends logarithmically
on $\rho$ and therefore corresponds to the na\"{\i}ve
solution of the 
$d=3$ reduced field equations. The $d=3$ fields read

\bdm
ds^2 = 
H^{-1} dt^2 - dx_i^2,\ \ \ e^{\phi}= H^{-1/2},
\edm

\noi
plus

\bdm
g_{\bar{k}\bar{k}}=H,\ \ \
\bar{\eta}_{\bar{k}9}=\frac{Q}{4 \pi R}\ \theta
\edm

\noi
in the electric and

\bdm
g_{\bar{k}\bar{k}}=H^{-1},\ \ \
\eta^{\bar{k}9}=\frac{P}{4 \pi R} \ \theta
\edm

\noi
in the magnetic case. This corresponds to the vortex solutions
studied in \cite{Sen95}.  Note that this solution is not 
asymptotically flat, therefore an
interpretation as soliton seems difficult \cite{Obe98}.

Under $\theta \rightarrow \theta + 2\pi$, we encounter 
the discrete $d=4$ charges by the shifts $\bar{\eta}_{\bar{k}9} 
\rightarrow \bar{\eta}_{\bar{k}9} + Q/2 R$ and
$\eta^{\bar{k}9}
\rightarrow \eta^{\bar{k}9} + P/2 R$. Full rotations therefore 
translate to discrete $E_{7(+7)}$ transformations 
parallel to \cite{Sen95}.

Note that the electric and magnetic solutions are embedded 
in a perfectly equal way. We actually expect the spectrum to 
significantly unify in $d=3$, since only particle and 
string-like configurations survive. 

This concludes our example of $d=4$ solitons in $d=3$.
The whole spectrum
in $d=3$ will be spanned by $E_{8(+8)}$ resp. a discrete subgroup
$E_{8(+8)}(\ZZ)$, which we will study now.

\subsection{The $E_{8(+8)}$ Coset in $d=3$}

We will recover the 
$E_{8(+8)}/SO(16)$ coset in this section. The decomposition of  
$E_{8(+8)}\supset SL(2,\RR) \times E_{7(+7)}$ in figure 
\ref{e8decomp} was already described in section \ref{chap}.
Again, the $SL(2,\RR)$ factor will carry the $d=3$ 
dilaton and dualized Kaluza-Klein gauge field.

Using (\ref{E8generatorrelations}) in appendix A, 
one may verify that the Lagrangian (\ref{e7}) takes the form

\bdm
{\cal L}
= - \frac14 E^{(3)} R^{(3)}
+\frac{1}{240}
E^{(3)}{\rm Tr}(P^{(3)}_{\mu}P^{(3) \mu})
\edm

\noi
with

\bdm
\partial_{\mu}{\cal V}^{(3)} {\cal V}^{(3) -1}= 
Q^{(3)}_{\mu} + P^{(3)}_{\mu},
\ \ Q^{(3)}_{\mu} \in \frak{e}_{8(+8)},
\ \ P^{(3)}_{\mu} \in \frak{e}_{8(+8)}-\frak{so}(16)
\edm

\noi
where 

\beq
{\cal V}^{(3)} = 
{\cal V}'^{(4)}
\exp\Big(\frac12\phi\ \sum_{i=1}^8 h_i \Big)
\exp\Big({{\cal Y}\cdot {\cal S}}\Big)
\exp\Big({f\ E^1_{\ 9}}\Big).
\label{3de7}
\eeq

\noi  
${\cal V}'^{(4)}$ is identical to (\ref{ve7}), but is now in 
the {\bf 248} adjoint representation of $E_{8(+8)}$.

The Lagrangain admits local $SO(16)$ and global $E_{8(+8)}$ symmetry.
Quite similar to the toy model, 
$E_{8(+8)}$ decomposes as

\beqa
E_{8(+8)}&\supset& SL(2,\RR)\times E_{7(+7)}
\n
{\bf 248}&\supset&({\bf 2},{\bf 56}) + ({\bf 1},{\bf 133}) + 
({\bf 3},{\bf 1}).
\label{e8dec}
\eeqa

\noi
${\cal V}'^{(4)}$ 
therefore has block structure

\bdm
{\cal V}'^{(4)}=\left[\begin{array}{cccccc}
\fbox{\bf 56}&&&&&\\&\fbox{$\overline
{\bf 56}$}&&&&\\&&\fbox{\rule[-0.4cm]{0cm}{1cm}
\bf 133}&&&\\&&&
\fbox{\bf 1}&&\\&&&&\fbox{\bf 1}&
\\&&&&&\fbox{\bf 1}\end{array}\right].
\edm

\subsection{Identifying $d=4$ U-Duality in the $d=3$ Theory}

To recover $d=4$ duality, consider an element

\bdm
\Lambda\in E_{7(+7)}, \ \ \
\Lambda = e^X, \ X\in \frak{e}_{7(+7)}
\edm

\noi
where $\frak{e}_{7(+7)}$ 
is generated by the set (\ref{e7generators}).
Parallel to (\ref{recover}) one gets 

\bdm
{\cal V}'^{(4)}\rightarrow {\cal V}'^{(4)} \Lambda^{-1} 
\ \mbox{and}\
{\cal Y}\rightarrow {\bf \rho_{56}}(\Lambda) {\cal Y}
\edm

\noi
and therefore recovers the $d=4$ U-duality. 

As we have seen, 
the discrete $d=4$ duality in $d=3$ corresponds to traveling
around vortex solutions on a full circle.  
The generators of the discrete group are identical to 
(\ref{discrete1}) resp. (\ref{discrete2}) in the {\bf 248} 
adjoint representation of $E_{8(+8)}$. 

\subsection{Connection to $d=11$ fields}

In order to compare different orders of compactification 
as indicated
in figure \ref{fige8}, we put the coset matrix 
${\cal V}^{(3)}$ in a convenient form.
Remembering the $d=11$ vielbein,

\beq
E_{~M}^{(11)A}=
\left[
\begin{array}{cc}e^{-1}E_{~\mu}^{(3)\alpha}
& B_{\mu}^{(3) i} e_i^a \\
\mbox{\Large 0}
&\makebox[2.0 cm]
{\rule[-0.7cm]{0cm}{1.5cm}$e_i^a$}
\end{array}
\right]
=
\left[
\begin{array}{cc}
e^{\phi/2}\Delta^{-\frac14}E_{~\mu}^{(3)\alpha}
& 
\begin{array}{cc}
e^{-\phi/2}\Delta^{-\frac14}\hat{B}_{\mu} & 
B^{(4) \bar{i}}_{\mu} \rho_{\bar{i}}^{\ \bar{a}}
\end{array} \\
\mbox{\Large 0}
&\parbox{4.5cm}{$
\rule[-0.5cm]{0cm}{1.3cm}
\begin{array}{cc}
\Delta^{-\frac14}e^{-\phi/2} & \ \ \ \ 
B^{(4) \bar{i}}_{3} \rho_{\bar{i}}^{\ \bar{a}}
\\
 0 &  \rho_{\bar{i}}^{\ \bar{a}} \end{array}$}
\end{array}
\right],
\label{elfbein}
\eeq

\noi
we define

\beqa
\varphi^{1\bar{i}}&=&\varphi^{(4)\bar{i}},
\n
\varphi^{\bar{i}\bar{j}}&=&
2\eta^{\bar{i}\bar{j}}
+4\varphi^{(4)[\bar{j}}\eta^{\bar{i}]9}
+\frac1{6}\epsilon^{\bar{i}\bar{j}
\bar{l}\bar{m}\bar{n}
\bar{p}\bar{q}}
A_{(\bar{l}+2)(\bar{m}+2)(\bar{n}+2)}
\bar{\eta}_{\bar{p}\bar{q}},
\n
\Psi_1&=&
f 
-\eta^{\bar{i}9}(
 -2\bar{\eta}_{\bar{i}9}
 +2\bar{\eta}_{\bar{i}\bar{j}}
  \varphi^{\bar{j}}
 -2\eta^{\bar{j}\bar{k}}
  A_{(\bar{i}+2)(\bar{j}+2)(\bar{k}+2)})\n
&&
-\frac1{36}\epsilon^{\bar{i}\bar{j}
\bar{k}\bar{l}
\bar{p}\bar{q}\bar{r}}
(\bar{\eta}_{\bar{i}\bar{j}}
-2\eta^{\bar{m}9}
A_{(\bar{i}+2)(\bar{j}+2)(\bar{m}+2)})
A_{(\bar{p}+2)(\bar{q}+2)(\bar{r}+2)}
\bar{\eta}_{\bar{k}\bar{l}}
\n
\Psi_{\bar{i}}&=&
-\bar{\eta}_{\bar{i}9}
+\bar{\eta}_{\bar{i}\bar{j}}
\varphi^{\bar{j}}
-A_{(\bar{i}+2)(\bar{j}+2)(\bar{k}+2)}
\eta^{\bar{j}\bar{k}}
-\frac1{36}\epsilon^{\bar{j}\bar{k}
\bar{l}\bar{m}\bar{n}
\bar{p}\bar{q}}
A_{(\bar{i}+2)(\bar{j}+2)(\bar{k}+2)}
A_{(\bar{l}+2)(\bar{m}+2)(\bar{n}+2)}
\bar{\eta}_{\bar{p}\bar{q}}.
\nonumber
\eeqa

\noi
Using (\ref{elfbein}) it may be checked that (\ref{3de7})
is identical to

\beqa
{\cal V}^{(3)} &=& 
\exp\left(\sum_{m=1}^{8} 
\ln\left(-\prod_{n=1}^m e_n^{\ \dot{n}}\right)\ h_m\right)
\prod_{p=0}^{6}
\exp \left(-\sum_{q,r=8-p}^8 e_{7-p}^{\ \ \ \ \ \dot{q}} 
(e^{(7-p)\ -1})^{\ \ r}_{\dot{q}}
\ E^{7-p}_{\ \ \ \  \ r} \right)
\n
&&
\exp\left(\sum_{i=1}^{8}\Psi_i\ 
E^i_{\ 9}\right)
\n
&&
\exp  
\left(\frac{2}{3!} \sum_{i,j,k=1}^8  
A_{(2+i)(2+j)(2+k)}\ E^{ijk} 
- \frac1{2!} \sum_{i,j=1}^8 \varphi^{ij}\ E^*_{ij9} 
\right)
\nonumber
\eeqa

\noi 
where summations have been spelled out. The 
$\varphi^{ij}$, $\Psi_i$ obey

\beqa
\partial_{\mu} \varphi^{ij} &=& 
-2 e^2 E^{(3)} \epsilon_{\mu\nu\rho} 
(\partial^{[\nu}A^{\rho](i+2)(j+2)} + 
B^{(3) k [\nu}\partial^{\rho]}A_{(k+2)(i+2)(j+2)})
\n
&&
+\frac{1}{18}\epsilon^{ijklmnpq} \partial_{\mu} 
A_{(2+k)(2+l)(2+m)} A_{(2+n)(2+p)(2+q)}
\n
\partial_{\mu} \Psi_i &=& 
-\frac12 e^2 E^{(3)} \epsilon_{\mu\nu\rho} B^{\nu\rho}_i 
-\frac12(\varphi^{kl} 
\partial_{\mu} A_{(2+k)(2+l)(2+i)} 
- \partial_{\mu}\varphi^{kl} A_{(2+k)(2+l)(2+i)}) 
\n
&&
-
\frac{1}{54}\epsilon^{jklmnpqr} 
 A_{(2+i)(2+j)(2+k)}
\partial_{\mu} A_{(2+l)(2+m)(2+n)} A_{(2+p)(2+q)(2+r)}
\nonumber
\eeqa

\noi
where $i,j,k,\dots =1\dots 8$.
This is exactly the result of \cite{Miz97} found by direct 
reduction to $d=3$.

\subsection{Different Orders of Compactification}

To study different orders of compactification, consider the 
internal vielbein $e_i^{\ a}$. Identifying the coordinate $2+n$
as fourth coordinate in four dimensions, 
as indicated in figure \ref{fige8}, corresponds to 
an exchange of the first and $n$th row of $e_i^{\ a}$. 
Ordering the
columns as $\{n, 2, 3,\dots,n-1,1,n+1, \dots, 8\}$, where the 
1 appears at the $n$th position, we need the new vielbein 
$\tilde{e}_i^{\ a}$ to be triangular. For this,    
one may always find an $SO(8)$ transformation such that the 
elfbein transforms like 

\beqa
&&
\left[
\begin{array}{cl}
e^{\phi/2}\Delta^{-\frac14}E_{~\mu}^{(3)\alpha}
& 
\begin{array}{cc}
e^{-\phi/2}\Delta^{-\frac14}\hat{B}_{\mu} & 
\ \ \; B^{(4) \bar{i}}_{\mu} \rho_{\bar{i}}^{\ \bar{a}}
\end{array} \\
\mbox{\LARGE 0}
&
\begin{array}{ccccccc}
\Delta^{-\frac14}e^{-\phi/2} & 
B^{(4) \bar{i}}_{3} \rho_{\bar{i}}^{\ \dot{2}}&
B^{(4) \bar{i}}_{3} \rho_{\bar{i}}^{\ \dot{3}}&&&\dots&
B^{(4) \bar{i}}_{3} \rho_{\bar{i}}^{\ \dot{8}}\\  
0& \rho_{2}^{\ \dot{2}}&&&&\dots &\rho_{2}^{\ \dot{8}}\\
0& 0 & \rho_{3}^{\ \dot{3}}&&& \dots & \rho_{3}^{\ \dot{8}}\\
&&&&&\dots&\\
0& \dots&&\ 0\ &\ \rho_{n}^{\ \dot{n}}\ & \dots& 
\rho_{n}^{\ \dot{8}}\\
&&&&&\dots&\\
0 &0&&& \dots  &0 &\rho_{8}^{\ \dot{8}}
\end{array}
\end{array}
\right]
\n
&\rightarrow&
\left[
\begin{array}{cl}
e^{\tilde{\phi}/2}\tilde{\Delta}^{-\frac14}E_{~\mu}^{(3)\alpha}
& 
\begin{array}{cc}
e^{-\tilde{\phi}/2}\tilde{\Delta}^{-\frac14}
\tilde{\hat{B}}_{\mu} & 
\ \ \; \tilde{B}^{(4) \bar{i}}_{\mu} 
\tilde{\rho}_{\bar{i}}^{\ \bar{a}}
\end{array} \\
\mbox{\LARGE 0}
&
\begin{array}{ccccccc}
0& \dots&&\ 0\ &\ \tilde{\rho}_{1}^{\ \dot{n}}\ & \dots& 
\tilde{\rho}_{1}^{\ \dot{8}}\\
0& \tilde{\rho}_{2}^{\ \dot{2}}&&&&\dots &\tilde{\rho}_{2}^{\ \dot{8}}\\
0& 0 & \tilde{\rho}_{3}^{\ \dot{3}}&&& \dots & \tilde{\rho}_{3}^{\ \dot{8}}\\
&&&&&\dots&\\
\tilde{\Delta}^{-\frac14}e^{-\tilde{\phi}/2} &
\tilde{B}^{(4) \bar{i}}_{2+n} \rho_{\bar{i}}^{\ \dot{2}}&
\tilde{B}^{(4) \bar{i}}_{2+n} \rho_{\bar{i}}^{\ \dot{3}}&&&\dots&
\tilde{B}^{(4) \bar{i}}_{2+n} \rho_{\bar{i}}^{\ \dot{8}}\\
&&&&&\dots&\\
0 &0&&& \dots  & 0 &\tilde{\rho}_{8}^{\ \dot{8}}
\end{array}
\end{array}
\right].\n
\label{local}
\eeqa
 
\noi 
Taking the sign change in the Chern-Simons term into account,
we may then write the scalar coset matrix for the 
compactification obtained by taking  
$\{0,1,2,(2+n)\}$ as four dimensional coordinates
as

\bdm
{\cal V}^{(3)}_{\# n} = 
{\cal V}^{(3)}_{\# 1} (\tilde{e}_i^{\ a}, 
\tilde{\Psi}_i,\tilde{A}_{(i+2),(j+2),(k+2)},
\tilde{\varphi}^{ij})
\edm

\noi
with 

\beqa
\tilde{A}_{({i}+2),({j}+2),({k}+2)}&=&
-{A}_{({i}+2),({j}+2),({k}+2)}, \ \
{i},{j},{k}\neq 1,n, \n
\tilde{A}_{3,({j}+2),({k}+2)}&=&
-{A}_{(n+2),({j}+2),({k}+2)}, \ \
{j},{k}\neq 1,n, \n
\tilde{A}_{(n+2),({j}+2),({k}+2)}&=&
-{A}_{3,({j}+2),({k}+2)}, \ \
{j},{k}\neq 1,n, \n
\tilde{A}_{3,(n+2),({k}+2)}&=&
-{A}_{(n+2),3,({k}+2)}, \ \
{k}\neq 1,n 
\nonumber
\eeqa

\noi
and

\beqa
\tilde{\varphi}^{ij}&=&
-\varphi^{ij}, \ \ i,j\neq 1,n,
\n 
\tilde{\varphi}^{1j}&=&
-\varphi^{nj}, \ \ j\neq 1,n,
\n 
\tilde{\varphi}^{nj}&=&
-\varphi^{1j}, \ \ j\neq 1,n,
\n 
\tilde{\varphi}^{1n}&=&
-\varphi^{n1},
\nonumber
\eeqa

\noi
finally

\beqa
\tilde{\Psi_i}&=&\Psi_i,\ \  i\neq1,n,
\n
\tilde{\Psi_1}&=&\Psi_n,
\n
\tilde{\Psi_n}&=&\Psi_1
\nonumber
\eeqa

\noi
while the U-duality group in this compactification 
is generated by 
(\ref{e7generators}) and the discrete group by
(\ref{discrete1}) resp. (\ref{discrete2}) for all
${\cal V}^{(3)}_{\# n}$.

\subsection{Joining U-dualities in $d=3$}

The above compactifications are related by

\bdm
{\cal V}^{(3)}_{\# n} = 
(P_n S^1_{\ n})^{-1} \; h_n\;\ 
{\cal V}^{(3)}_{\# 1}\;\ P_n S^1_{\ n} 
\edm

\noi
where $h_n$ is the natural lift to $E_{8(+8)}$ 
of the local transformation
(\ref{local}),
$S^1_{\ n}$ generates the  
Weyl reflection exchanging $\eps_1$ with 
$\eps_n$ and $P_n$ is again a ``parity'' transformation. One has

\beqa
P_n&=&(-1)^{h_{678} + h_6 + \sum\limits_{m=n}^7 h_m} = 
(S^{678})^2 (S^6_{\ 7})^2 (S^n_{\ 8})^2\ \ \ 2\leq n \leq 7,\n
P_8&=&(-1)^{h_{678} + h_6}= (S^{678})^2 (S^6_{\ 7})^2.
\nonumber
\eeqa

\noi
$P_n$ is an $E_{7(+7)}(\ZZ)$ transformation corresponding to a charge 
conjugation in $d=4$, but leaves the fields 
$\tilde{A}^{n\bar{j}}_{\bar{\mu}\bar{\nu}}$, 
$B^{(4)\bar{j}}_{\bar{\mu}\bar{\nu}}$, $\bar{j}\neq n$ unchanged.
Note that for $n=8$ they exactly correspond to the NS-NS fields in
section 3.2.

The $d=3$ U-duality is therefore given by joining all

\bdm
\Lambda_n= P_n S^1_{\ n} \ \Lambda\  (P_n S^1_{\ n})^{-1}
\edm

\noi  
where $\Lambda$ is an element of $E_{7(+7)}$ spanned by  
(\ref{e7generators}), and in the discrete 
case by (\ref{discrete1}) resp. (\ref{discrete2}).

Note that the intersection of two different U-dualities is 
exactly $E_{6(+6)}$ as expected!
Joining all $\Lambda_n$ 
gives the whole of $E_{8(+8)}$, and the $d=3$ discrete
U-duality is generated by the set of generators obtained
by exponentiating the Chevalley generators for all roots. This 
coincides with $E_{8(+8)}(\ZZ)$ as defined in appendix B.

We have therefore given a complete set of 
generators for the U-duality group in 
three dimensions.

\subsection{$G_{2(+2)}$ in $E_{8(+8)}$}

Before concluding, we would like to turn back to our toy model.

$\frak{g}_{2(+2)}$ may be found in the algebra of 
$\frak{e}_{8(+8)}$ by considering the direct embedding 
$\frak{sl}(9) \supset \frak{sl}(3)$. We may choose for example  

\bdm
E^{i12} +E^{i34} +E^{i56}, E^*_{i12} +E^*_{i34} +E^*_{i56},
E_i^j (i,j=7,8,9).
\edm

It may be checked by using (\ref{E8generatorrelations})
that this choice generates $\frak{g}_{2(+2)}$.
From the point of view of the physical theory, 
this truncation is suggested by identifying 
(\{0,1,2,3,4\} as $d=5$ coordinates)

\beqa
&ds^{(11)\; 2} = ds^{(5)\; 2} +  ds^{(E6)\; 2}&\n
&A^{(11)} = -\frac{1}{\sqrt{3}} A^{(5)} \wedge J ,\ 
J=\frac{1}{2}(
dx^5 \wedge dx^6 + dx^7 \wedge dx^8 + dx^9 \wedge dx^{10})&
\eeqa

\noi
where $E6$ is the flat six dimensional Euclidean space and 
$J$ is its K\"ahler form, as proposed in
\cite{Pap96}, and $A^{(11)}$, $A^{(5)}$ is the 
eleven dimensional three form and the five dimensional one-form 
potential. 

Note that reexpressing the $d=5$ one form by its M-theory 
pendant exactly cancels the factor $\frac13$ in 
(\ref{DSZshifted}). The factor $3$ is actually symptomatic 
for the truncation $E_{8(+8)}$ to $G_{2(+2)}$, e.g the length 
squared of the long roots is 3 times larger than the one 
of the short roots.  

In the toy model, we found no agreement of the discrete 
U-duality group with the definition of appendix B, while
there was agreement for M-theory. However, this may be seen 
as a consequence of the rather complicated embedding of 
$\frak{g}_{2(+2)}$ into $\frak{e}_{8(+8)}$, involving 
generator sums at the level of the algebra, and leading to a
non-simply laced algebra.
   
\section{Conclusions and Outlook}

In this paper, we have studied the discrete U-duality groups 
in the context of low energy supergravity as proposed by Hull 
and Townsend \cite{Hul95}. We have proposed a set of generators 
for $E_{7(+7)}$, corresponding to the $d=4$ U-duality group, 
and presented a proof. Higher dimensional U-dualities are 
found by direct embedding. 

In studying this group, we have made the {\bf 56} 
representation of
$E_{7(+7)}$ explicit by an embedding into $E_{8(+8)}$. 
The way the representation was given might 
make it ``manageable'' in order
to investigate U-duality 
transformations connecting BPS solutions as
mentioned in \cite{Ber99}.

In comparison to the other definitions of the U-duality group, we 
have seen that they agree with our definition and that 
the set of generators used in the algebraic approach to 
U-duality can be identified with ours. 

We have extended a proposal by Hull and Townsend
along the lines of Sen to determine 
the $d=3$ U-duality group and found that this indeed yields 
the full $E_{8(+8)}(\ZZ)$ in the definition of appendix B
and \cite{Matsumoto}.
It is interesting to note that the transition from one 
compactification to the other involves a ``parity'' transformation,
corresponding to a charge conjugation in $d=4$. The fact that we
obtained a proposed $d=3$ quantized symmetry from a $d=4$ quantization
condition is interesting in this context and might yield new insights
into the quantization in $d=3$.

We have studied this procedure also in a truncated model 
corresponding  to simple $d=5$ supergravity, and 
seen that the $d=3$ U-duality group in this case {\it fails} 
to agree
with the definition of appendix B and \cite{Matsumoto}, 
that is, the U-duality 
in $d=3$ is strictly smaller. 
We conclude that 
truncations of M-theory  
need to be studied very carefully, especially in a quantized 
context.

The procedure we have presented to determine low dimensional 
discrete U-duality groups can be 
extended to $d=2$ and even $d=1$. 
In these theories, a classical Ka\v{c}-Moody symmetry 
algebra known 
as $\frak{e}_9$ and Yangian algebra $Y(\frak{e}_8)$ upon quantization
(see \cite {Koe99} for a recent account) is present.
By a lightlike reduction, which makes 
an intriguing connection to the infinite momentum frame 
of Matrix theory,  
a hyperbolic Ka\v{c}-Moody algebra denoted by $\frak{e}_{10}$ is 
expected (\cite{Miz97}, see \cite{E10} for an introduction to
hyperbolic Ka\v{c}-Moody algebras).
In this context, it is interesting to note that 
the notion of discrete groups over $\ZZ$ has been connected to 
homeomorphisms of the group ring over $\ZZ$ to $\ZZ$
\cite{Chevalley, Kostant, Matsumoto}, thereby 
making a connection to Hopf algebra structures. 
It should be interesting to see how the notion of discrete 
duality groups extends for $d=2$ and $d=1$. 
Work in this direction is in progress.

\bigskip
\noi
{\large\bf Acknowledgements}\\
We would like to thank Hermann
Nicolai for reading the manuscript and suggestions.
S.M. also thanks the Albert-Einstein-Institute 
Potsdam for hospitality, where a part of this work was done.

\appendix
\section{Exceptional Lie Algebras}

In this appendix, we will present the representation of 
exceptional Lie algebras parallel to Freudenthal's
paper \cite{Freudenthal}. We will start with 
$\frak{g}_{2(+2)}$, and 
then turn to $\frak{e}_{8(+8)}$.

\subsection{Realization of $\frak{g}_2$}
The Lie algebra $\frak{g}_2$ is known to 
allow a $\ZZ_3$ grading and 
decomposes into the adjoint and two fundamental 
representations of 
$\frak{sl}(3)$:
\begin{eqnarray}
{\bf 14} = {\bf 8}\oplus{\bf 3}\oplus\overline{\bf 3}.
\end{eqnarray}
Thus any element of $\frak{g}_2$ can be 
specified by a traceless matrix 
$X_{i'}^{~j'}$, $1\leq i',j' \leq 3$ and two vectors 
$v_{i'}$ and $v^{*i'}$, $1\leq i'\leq 3$ which transform as 
${\bf 3}$ and $\overline{\bf 3}$ under $\frak{sl}(3)$.
The Lie bracket of the two elements 
$[X_{(1)},v_{(1)},v^*_{(1)}]$,
$[X_{(2)},v_{(2)},v^*_{(2)}]$ gives rise to a new element of 
$\frak{g}_2$ specified by $[X_{(3)},v_{(3)},v^*_{(3)}]$ with 

\begin{eqnarray}
X_{(3)i'}^{~~~~j'}&=&X_{(1)i'}^{~~~~l'}X_{(2)l'}^{~~~~j'}
-X_{(2)i'}^{~~~~l'}X_{(1)l'}^{~~~~j'}\nonumber\\
&&-3 \left(v_{(1)i'}v_{(2)}^{*j'}
                  -v_{(2)i'}v_{(1)}^{*j'}
\right.\nonumber\\
&&~~~~~-\frac13\left.(v_{(1)p'}v_{(2)}^{*p'}
                     -v_{(2)p'}v_{(1)}^{*p'})\delta_{i'}^{j'}
\right)
\nonumber\\
v_{(3)i'}&=&X_{(1)i'}^{~~~~~l'}v_{(2)l'}
-X_{(2)i'}^{~~~~~l'}v_{(1)l'}\nonumber\\
&&-2\epsilon_{i'j'k'}
v^{*j'}_{(1)}v^{*k'}_{(2)}\nonumber\\
v_{(3)}^{*i'}&=&-(X_{(1)l'}^{~~~~i'}v_{(2)}^{*l'}
-X_{(2)l'}^{~~~~i'}v_{(1)}^{*l'})\nonumber\\
&&+2\epsilon^{i'j'k'}
v_{(1)j'}v_{(2)k'},\label{LiebracketG2}
\end{eqnarray}
where $\epsilon_{i'j'k'}
=-\epsilon^{i'j'k'}$ \footnote{
The indices are raised and lowered by the metric
$\mbox{diag}[-1,-1,-1]$.}.
If all 
$X_{i'}^{~j'}$, $v_{i'}$ and $v^{*i'}$ are restricted to
real numbers, the relations (\ref{LiebracketG2}) define
the real form $\frak{g}_{2(+2)}$.

The roots of $\frak{g}_2$ are shown in Figure 3.  
Regarded as weights of $\frak{sl}(3)$, 
they are naturally embedded 
in a hyperplane in ${\bf R^3}$: 

\begin{eqnarray}
\eps_{ij}&\equiv&\mbox{\boldmath $\epsilon_i$}-
                 \mbox{\boldmath $\epsilon_j$}
\hskip 5ex (1\leq i\neq j\leq 3),\nonumber\\
\pm\eps_{i}&\equiv&
\pm (\mbox{\boldmath $\epsilon_i$}
-\frac13\sum_{l=1}^3\mbox{\boldmath $\epsilon_l$})
\hskip 5ex (1\leq i\leq 3),
\end{eqnarray}
where $\{\mbox{\boldmath $\epsilon_i$}~|~1\leq i\leq 3\}$ 
is a set of orthonormal
vectors in ${\bf R^3}$. The hyperplane is normal to
$\sum_{i=1}^3\mbox{\boldmath $\epsilon_i$}$.

\begin{figure}[htbp]
\begin{center}
\leavevmode
\begin{picture}(0,0)%
\epsfig{file=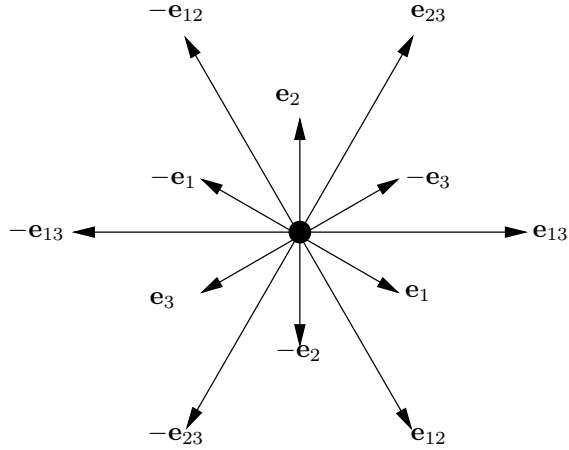}%
\end{picture}%
\setlength{\unitlength}{4144sp}%
\begin{picture}(3240,2517)(1081,-2887)
\put(4321,-1681){\makebox(0,0)[lb]{\footnotesize $\eps_{13}$}}
\put(3556,-1366){\makebox(0,0)[lb]{\footnotesize $-\eps_{3}$}}
\put(3591,-366){\makebox(0,0)[lb]{\footnotesize $\eps_{23}$}}
\put(2781,-871){\makebox(0,0)[lb]{\footnotesize $\eps_{2}$}}
\put(2016,-366){\makebox(0,0)[lb]{\footnotesize $-\eps_{12}$}}
\put(2031,-1366){\makebox(0,0)[lb]{\footnotesize $-\eps_{1}$}}
\put(1181,-1681){\makebox(0,0)[lb]{\footnotesize $-\eps_{13}$}}
\put(2031,-2086){\makebox(0,0)[lb]{\footnotesize $\eps_{3}$}}
\put(2016,-2900){\makebox(0,0)[lb]{\footnotesize $-\eps_{23}$}}
\put(2781,-2401){\makebox(0,0)[lb]{\footnotesize $-\eps_{2}$}}
\put(3591,-2900){\makebox(0,0)[lb]{\footnotesize $\eps_{12}$}}
\put(3556,-2041){\makebox(0,0)[lb]{\footnotesize $\eps_{1}$}}
\end{picture}
\end{center}
\caption{\label{g2roots} Roots of $\frak{g}_{2(+2)}$}
\end{figure}

Let us write 
$E_{\eps_{ij}}=E^i_{\;\;j}$,
$E_{\eps_i}=E^i$ and 
$E_{-\eps_i}=E^*_i$
corresponding to each non-zero root, respectively. We also take 
$\{h_i \equiv [E^i_{\;\;i+1},E^{i+1}_{~~~~i}]\;|\; i=1,2 \}$
as the basis of the Cartan subalgebra. Their commutators are: 

\begin{eqnarray}
{[}~h_i ~,~h_j ~{]}&=&0, \nonumber\\
{[}~h_i ~,~E^j_{\;\;k} ~{]}&=&
\delta_i^j E^i_{\;\;k} - \delta_{i+1}^j E^{i+1}_{~~~~k}
-\delta_k^i E^j_{\;\;i} + \delta_k^{i+1} E^j_{\;\;i+1}, 
\nonumber\\
{[}~h_i ~,~E^j ~{]}&=& \delta_i^j E^i
                     -\delta_{i+1}^j E^{i+1},\nonumber\\
{[}~h_i ~,~E^*_j ~{]}&=& -(\delta_j^i E^*_i
                     -\delta_j^{i+1}E^*_{i+1}),\nonumber\\
{[}~E^i_{\;\;j} ~,~E^k_{\;\;\;l} ~{]}&=&
                     \delta^k_j E^i_{\;\;l}-\delta^i_l 
E^k_{\;\;j}, \nonumber\\
{[}~E^i_{\;\;j} ~,~E^k ~{]}&=&
                     \delta^k_j E^i, \nonumber\\
{[}~E^i_{\;\;j} ~,~E^*_k ~{]}&=&
                     -\delta^i_k E^*_j, \nonumber\\
{[}~E^i ,E^j ~{]}&=&
                     -2\sum_k^3 \epsilon^{ijk}
                     E^*_k, \nonumber\\
{[}~E^*_i ,E^*_j ~{]}&=&
                     -2\sum_k^3\epsilon_{ijk}
                     E^k, \nonumber\\
{[}~E^i ,E^*_j ~{]}&=&
                     3 E^i_{\;\;j}
                     \hskip 2ex\mbox{if $i\neq j$},\nonumber\\
{[}~ E^1 ,E^*_1 ~{]}&=&2 h_1+ h_2,\nonumber\\
{[}~ E^2 ,E^*_2 ~{]}&=&- h_1+ h_2,\nonumber\\
{[}~ E^3 ,E^*_3 ~{]}&=&- h_1-2 h_2.
\label{G2generatorrelations}
\end{eqnarray}

\noi
Their expressions in terms of the
triple are
\begin{eqnarray}
h_i&=&[\delta_{i'}^i \delta_i^{j'}
      -\delta_{i'}^{i+1} \delta_{i+1}^{j'},0,0],
\nonumber\\
E^i_{\;\;j}&=&[\delta_{i'}^i \delta_j^{j'},0,0],
\nonumber\\
E^{i}&=&[0,\delta_{i'}^i,0],
\nonumber\\
E^*_{i}&=&[0,0,-\delta_i^{i'}].
\end{eqnarray}

\subsection{Realization of $E_8$}
The adjoint representation of $E_8$ is decomposed into a sum
of irreducible representations of the subalgebra $SL(9)$ as
\begin{eqnarray}
{\bf 248} = {\bf 80}\oplus{\bf 84}\oplus\overline{\bf 84},
\end{eqnarray}
where {\bf 80} is the defining representation, while
{\bf 84} ($\overline{\bf 84}$) is (3,0) ((0,3)) antisymmetric
tensor representation. One may specify any element of
$E_8$ by a triple:
\begin{eqnarray}
                  [X_{i'}^{~j'},~v_{i'j'k'},~v^{*i'j'k'}]
                  \hskip 2ex
                  (\mbox{$X$ is traceless;
                    $v$,$v^*$ are totally antisymmetric}),
\end{eqnarray}
where $i',j',\ldots=1,\ldots,9$ are the indices of the
nine dimensional vector space on which $X$, $v$ and $v^*$ act.
The Lie bracket of the two elements $[X_{(1)},v_{(1)},v^*_{(1)}]$
and $[X_{(2)},v_{(2)},v^*_{(2)}]$ is again a triple
$[X_{(3)},v_{(3)},v^*_{(3)}]$ with \cite{Freudenthal}
(Again, the repeated indices are summed over.)
\begin{eqnarray}
X_{(3)i'}^{~~~~j'}&=&X_{(1)i'}^{~~~~l'}X_{(2)l'}^{~~~~j'}
-X_{(2)i'}^{~~~~l'}X_{(1)l'}^{~~~~j'}\nonumber\\
&&-\frac1{2!}\left(v_{(1)i'p'q'}v_{(2)}^{*j'p'q'}
                  -v_{(2)i'p'q'}v_{(1)}^{*j'p'q'}
\right.\nonumber\\
&&~~~~~-\frac19\left.(v_{(1)p'q'r'}v_{(2)}^{*p'q'r'}
                     -v_{(2)p'q'r'}v_{(1)}^{*p'q'r'})
\delta_{i'}^{j'}\right)
\nonumber\\
v_{(3)i'j'k'}&=&3(X_{(1)[i'}^{~~~~~l'}v_{(2)j'k']\;l'}
-X_{(2)[i'}^{~~~~~l'}v_{(1)j'k']\;l'})\nonumber\\
&&-\frac1{3!}\epsilon_{i'j'k'l'm'n'p'q'r'}
v^{*l'm'n'}_{(1)}v^{*p'q'r'}_{(2)}\nonumber\\
v_{(3)}^{*i'j'k'}&=&-3(X_{(1)l'}^{~~~~[i'}v_{(2)}^{*j'k']\;l'}
-X_{(2)l'}^{~~~~[i'}v_{(1)}^{*j'k']\;l'})\nonumber\\
&&+\frac1{3!}\epsilon^{i'j'k'l'm'n'p'q'r'}
v_{(1)l'm'n'}v_{(2)p'q'r'},\label{Liebracket}
\end{eqnarray}
where $\epsilon_{i'j'k'l'm'n'p'q'r'}
=-\epsilon^{i'j'k'l'm'n'p'q'r'}$ \footnote{
The indices are raised and lowered by the metric
$\mbox{diag}[-1,\ldots,-1]$.}.
If all the tensors
$X_{i'}^{~j'}$, $v_{i'j'k'}$ and $v^{*i'j'k'}$ are restricted to
real numbers, then the relations (\ref{Liebracket}) define
the real form $E_{8(+8)}$.

Let us think of the eight dimensional root space of $E_8$
as a hyperplane lying in ${\bf R^9}$. Our non-zero roots are 
\begin{eqnarray}
\eps_{ij}&\equiv&\eps_i-\eps_j
\hskip 5ex (1\leq i\neq j\leq 9),\nonumber\\
\pm\eps_{ijk}&\equiv&
\pm (\eps_i+\eps_j+\eps_k
-\frac13\sum_{l=1}^9\eps_l)
\hskip 5ex (1\leq i<j<k\leq 9),
\end{eqnarray}
where $\{\eps_i~|~1\leq i\leq 9\}$ is a set of orthonormal
vectors in ${\bf R^9}$. The hyperplane is normal to
$\sum_{i=1}^9\eps_i$.

We write the corresponding generators as
\begin{eqnarray}
&&E_{\eps_{ij}}=E^i_{\;\;j}\hskip 2ex\mbox{(total 72)},
\nonumber \\
&&E_{\eps_{ijk}}=E^{ijk}\hskip 2ex\mbox{(total 84)},\nonumber \\
&&E_{-\eps_{ijk}}=E^*_{ijk}\hskip 2ex\mbox{(total 84)}
\end{eqnarray}
and take the commutators: 
\begin{eqnarray}
h_i\equiv [E^i_{\;\;i+1},E^{i+1}_{~~~~i}]\hskip 2ex
\mbox{(total 8)}
\end{eqnarray}
as the basis of the Cartan subalgebra $\{h_i \;|\; i=1,...,8 \}$.
These 72+84+84+8
= 248 generators form a basis of $E_8$ and
satisfy the relations (Note that the repeated indices 
are {\it not}
summed over unless stated explicitly.)
\begin{eqnarray}
{[}~h_i ~,~h_j ~{]}&=&0, \nonumber\\
{[}~h_i ~,~E^j_{\;\;k} ~{]}&=&
\delta_i^j E^i_{\;\;k} - \delta_{i+1}^j E^{i+1}_{~~~~k}
-\delta_k^i E^j_{\;\;i} + \delta_k^{i+1} E^j_{\;\;i+1}, 
\nonumber\\
{[}~h_i ~,~E^{jkl} ~{]}&=& 3(\delta_i^{[l}E^{jk]i}
                     -\delta_{i+1}^{[l}E^{jk]\;i+1}),\nonumber\\
{[}~h_i ~,~E^*_{jkl} ~{]}&=& -3(\delta_{[l}^iE^*_{jk]i}
                     -\delta_{[l}^{i+1}E^*_{jk]\;i+1}),
\nonumber\\
{[}~E^i_{\;\;j} ~,~E^k_{\;\;\;l} ~{]}&=&
                     \delta^k_j E^i_{\;\;l}-\delta^i_l 
E^k_{\;\;j}, \nonumber\\
{[}~E^i_{\;\;j} ~,~E^{klm} ~{]}&=&
                     3\delta^{[m}_j E^{kl]i}, \nonumber\\
{[}~E^i_{\;\;j} ~,~E^*_{klm} ~{]}&=&
                     -3\delta^i_{[m} E^*_{kl]j}, \nonumber\\
{[}~E^{ijk} ,E^{lmn} ~{]}&=&
                     -\frac1{3!}\sum_{p,q,r}^9 
\epsilon^{ijklmnpqr}
                     E^*_{pqr}, \nonumber\\
{[}~E^*_{ijk} ,E^*_{lmn} ~{]}&=&
                     -\frac1{3!}\sum_{p,q,r}^9
\epsilon_{ijklmnpqr}
                     E^{pqr}, \nonumber\\
{[}~E^{ijk} ,E^*_{lmn} ~{]}&=&
                     6\delta^j_{[m}\delta^k_n E^{i}_{\;\;l]}
                     \hskip 2ex\mbox{if $i\neq l,m,n$},
\nonumber\\
{[}~ E^{ijk} ,E^*_{ijk} ~{]}&=&-\frac13\sum_{l=1}^8 l h_l
                     +\sum_{l=i}^8 h_l
                     +\sum_{l=j}^8 h_l
                     +\sum_{l=k}^8 h_l\nonumber\\
                     &\equiv&h_{ijk}.
\label{E8generatorrelations}
\end{eqnarray}
We also extended the definition of $E^{ijk}$ as
\begin{eqnarray}
E^{ijk}=E^{jki}=E^{kij}=-E^{ikj}=-E^{jik}=-E^{kji},
\end{eqnarray}
and likewise for $E^*_{ijk}$. Their expressions in terms of the
triple are
\begin{eqnarray}
h_i&=&[\delta_{i'}^i \delta_i^{j'}
      -\delta_{i'}^{i+1} \delta_{i+1}^{j'},0,0],
\nonumber\\
E^i_{\;\;j}&=&[\delta_{i'}^i \delta_j^{j'},0,0],
\nonumber\\
E^{ijk}&=&[0,3!\delta_{[i'}^i\delta_{j'}^j\delta_{k']}^k,0],
\nonumber\\
E^*_{ijk}&=&[0,0,-3!\delta_i^{[i'}\delta_j^{j'}\delta_k^{k']}].
\end{eqnarray}
Also
\begin{eqnarray}
h_{ijk}=[\delta_{i'}^i \delta_i^{j'}
+\delta_{i'}^j \delta_j^{j'}
+\delta_{i'}^k \delta_k^{j'}
-\frac13\delta_{i'}^{j'},0,0
].
\end{eqnarray}
Thus $h_i$, $h_{ijk}$ may be thought of as
$E^i_{\;\;i}-E^{i+1}_{~~\;\;i+1}$, 
$E^i_{\;\;i}+E^{k}_{\;\;k}+E^{k}_{\;\;k}
-\frac13 \sum_{l=1}^9 E^l_{\;\;l}$, respectively.

Finally, we give some useful trace formulas:

\beqa
\mbox{Tr}_{248} E^i_{~j} E^k_{~l} &=& 60 \delta^i_l\delta^k_j,\n
\mbox{Tr}_{248} E^{ijk} E^*_{lmn} &=& 
                60\cdot 3!\delta^{[i}_l\delta^j_m\delta^{k]}_l,
\eeqa
$i,j,k=1,2,\ldots,9$.
\vskip 2em

\section{Discrete Subgroups of Lie Groups}

In this appendix we study the properties of the discrete 
subgroup 
$G(\ZZ)$ used in this paper. We give a definition of $G(\ZZ)$ 
and 
propose a set of generators. We use properties of a special 
class 
of representations relevant to dualities, together with the 
Birkhoff 
decomposition of Lie groups \cite{MP} to prove that all 
elements of 
$G(\ZZ)$ can be generated from this set.

Let $G$ be a complex simple Lie group, and $\g$ be the Lie 
algebra 
of $G$. Let $h$ be a Cartan subalgebra of $\g$, and $H$ be 
its Lie 
group. Let $\Phi$ be the set of roots of $G$ relative to $H$, 
$\Delta$ a fixed set of simple roots in $\Phi$ and 
$\Phi^+(\Phi^-)$ 
the set of positive(negative) roots
with respect to $\Delta$.

Consider a nontrivial irreducible representation $\rho$ of $G$ 
and let $\Lambda_\rho$ be the set of weights of $\rho$.
Let us call $\rho$ a {\it minimal} representation if all 
$\lambda\in \Lambda_\rho$ are transformed into each other by the 
Weyl group ${\cal W}$ of $G$. It is easy to show that $\rho$ 
is a 
minimal representation if and only if $\langle\lambda,~\alpha\rangle
(=2(\lambda,~\alpha)/(\alpha,~\alpha))=0$ 
or $\pm 1$ for all $\lambda\in\Lambda_\rho$, $\alpha\in\Phi$. 
Each coset of the root lattice $\Lambda_r$ in the weight lattice 
$\Lambda$ contains precisely one highest weight of a minimal 
representation. Thus there are $|\Lambda/\Lambda_r|-1$ such 
representations. In fact they are all fundamental 
representations. 
(See \cite{Humphreys}, page 72.)
It is enough to consider only this particular class of 
representations,
since they include all the representations relevant to our 
discussion 
such as the {\bf 56} of $E_7$, the 
$\mbox{\boldmath $N$}$ of $SL(N)$ and the {\bf 2}
\mbox{\boldmath $N$} 
of $SO(2N)$ \footnote{In Ref.\cite{Matsumoto} a larger class of 
representations which may include zero weights are considered 
and 
called {\it basic} representations. It is also claimed that 
$G(\ZZsub)$ 
does not depend on the representation.}.

Let $V$ be a representation space of a minimal representation 
$\rho$. 
A {\it lattice} $V_{\ZZsub}$ in $V$ is defined to be the 
$\ZZ$-span of 
a basis of $V$. With a fixed Chevalley basis 
$\{e_\alpha,\alpha\in\Phi;~ 
h_i, 1\leq i\leq N\}$  of $\g$, a lattice $V_{\ZZsub}$ 
is said to be 
{\it admissible} if $V_{\ZZsub}$ is invariant under the 
action of the 
$\ZZ$-form ${\cal U}(\g)_{\ZZsub}$ \cite{Humphreys} 
of the universal enveloping 
algebra ${\cal U}(\g)$, that is, invariant under the 
actions of all $\rho(e_\alpha)^n/n!$, $n\in\NN$. 
The discrete group $G(\ZZ)$ of $G$ is defined as the 
subgroup of $G$ 
which consists of all $g\in G$ such that $\rho(g)$ stabilizes 
$V_{\ZZsub}$ \cite{Kostant}.

\noindent
{\it Remark.} An admissible lattice fixes in which basis of $V$ 
the entries of the elements of $G(\ZZ)$ are restricted 
to integer 
values. For example, 
\[
SL(2,\ZZ)=\left\{
\left[\begin{array}{cc}a&b\\c&d\end{array}
\right]~|~a,b,c,d\in\ZZ,~ad-bc=1
\right\}
\]
stabilizes the admissible lattice
\[
\ZZ\left[\begin{array}{c}1\\0\end{array}
\right]
\oplus
\ZZ\left[\begin{array}{c}0\\1\end{array}
\right].
\]
In other generic basis of $V$, $SL(2,\ZZ)$ is not represented as 
a group of matrices with integral entries.
In $d=4$ duality in general, each $U(1)$ 
gauge field corresponds to different weight space and 
one can always 
normalize the charge lattice so that it may coincide with an 
admissible lattice. Therefore this definition of $G(\ZZ)$ and 
the concept of discrete duality groups agree. 
\vskip 1ex

One can construct the smallest admissible lattice $V_{\ZZsub}$ 
by acting 
${\cal U}(\g)_{\ZZsub}$ to a given highest weight 
vector of $V$. 

We will now prove the following proposition:
\vskip 1ex

\noindent {\bf Proposition \cite{Matsumoto}.} 
$G(\ZZ)$ coincides with the group generated by 
\[
\{\exp e_\alpha | \alpha\in\Phi\}
\]
denoted by $E(\ZZ)$ in the following. 
\vskip 1ex

\noindent
{\it Example.}
As is well known, $SL(2,\ZZ)$ is generated by 
\[
\exp e_\alpha = \left[
\begin{array}{cc}1&1\\0&1\end{array}
\right]~~~\mbox{and}~~~
\exp e_{-\alpha} = \left[
\begin{array}{cc}1&0\\1&1\end{array}
\right],
\]
or equivalently $T=\exp e_\alpha$ and 
$S=\exp e_\alpha \exp(- e_{-\alpha}) \exp e_\alpha$.
\vskip 1ex

\noindent
{\it Proof of Proposition.} Let $\alpha_1$ be the simple 
root dual 
to the highest weight $\lambda_1$ of $\rho$. 
Let ${\Phi'}^+$ be the set of positive roots orthogonal to 
$\lambda_1$:  
${\Phi'}^+=\{\alpha\in\Phi^+|(\lambda_1,\alpha)=0\}$. 
We introduce and fix an order (denoted by $<$) among the 
elements of 
$\Phi^+$ in such a way that $\alpha<\beta$ if 
$\alpha\in\Phi^{'+}$ 
and $\beta\in\Phi^+-\Phi^{'+}$. Similarly, for 
$-\alpha,-\beta \in\Phi^-$,  
we define $-\beta<-\alpha$ if $\alpha<\beta$.

We can take a basis $\{v_\lambda|\lambda\in\Lambda_\rho\}$
of $V_{\ZZsub}$ as
\beqa
v_{\lambda_1},&&\n
v_{\lambda_1-\alpha}&=&\rho(e_{-\alpha})v_{\lambda_1}
~~~(\alpha\in\Phi^+-{\Phi'}^+),\n 
v_{\lambda_1-\alpha-\beta}
&=&\rho(e_{-\beta})\rho(e_{-\alpha})v_{\lambda_1}
~~~(\alpha,\beta\in\Phi^+-{\Phi'}^+),\n 
\cdots&& \nonumber
\eeqa
where the product of $\rho(e_{-\alpha})$'s are 
``normal-ordered'' 
according to the order introduced above, that is, 
$\rho(e_{-\beta})$ 
comes to the left of $\rho(e_{-\alpha})$ if $-\beta<-\alpha$.

It is known that any $g\in G$ can be written 
in the form (the Birkhoff decomposition) 
\beqa
g&=&\prod_{\alpha\in\Phi^+}\exp(c_{-\alpha} e_{-\alpha})
\cdot w(g) \cdot 
\prod_{\alpha\in\Phi^+}\exp(c_\alpha e_\alpha)
\label{Birkhoff}
\eeqa
with $w(g)$ being an element of the normalizer of $H$.  
The multiple products are ``normal-ordered'' similarly.  
Since $\rho$ is a minimal representation, 
$\rho(e_\alpha)^2=0$ for any 
root $\alpha$. Thus each factor 
$\rho(\exp c_\alpha e_\alpha)$ in 
$\rho(g)$ (\ref{Birkhoff}) can be replaced by 
$1+ c_\alpha \rho(e_\alpha)$.

Let us write 
\beq
\rho(g)v_{\lambda_1}
=\sum_{\lambda\in\Lambda_\rho}
\rho(g)_{\lambda\lambda_1}v_\lambda,~~~
\rho(g)_{\lambda\lambda_1}\in\ZZ. \label{BwB}
\eeq 
Let $s_\alpha=\exp(-e_\alpha) \exp e_{-\alpha} \exp (-e_\alpha) 
\in E_{\ZZsub}$, $\alpha\in\Phi$. Then $\rho(s_\alpha)$ sends 
$v_\lambda$ to a vector proportional to 
$v_{\sigma_\alpha(\lambda)}$, 
where $\sigma_\alpha$ is the Weyl reflection with respect to 
$\alpha$. Since any weight of $\rho$ is transformed 
to the highest 
weight by the Weyl group, one can find some $s\in E_{\ZZsub}$ 
(written as a product of $s_\alpha$'s) such that 
$\rho(s)v_\lambda\propto v_{\lambda_1}$ for any weight 
$\lambda$. 
Therefore, if $\rho(g)_{\lambda_1\lambda_1}=0$, one may 
still have 
$\rho(sg)_{\lambda_1\lambda_1}\neq 0$ for some element $s$ of 
$E_{\ZZsub}$. Thus we assume that 
$\rho(g)_{\lambda_1\lambda_1}\neq 0$ 
from the beginning, without loss of generality.

Since $\rho(e_\alpha)v_{\lambda_1}=0$ for any positive 
root $\alpha$, and since $\rho(e_\alpha)_{\lambda_1\lambda}
\neq 0$ 
only if $\lambda=\lambda_1$ for any negative root $\alpha$ and 
any weight $\lambda$, our assumption  
$\rho(g)_{\lambda_1\lambda_1}\neq 0$ implies that $w(g)$ in 
(\ref{BwB}) 
stabilizes the highest weight $\lambda_1$, i.e. 
$\rho(w(g))v_{\lambda_1}=c v_{\lambda_1}$ for some constant 
$c\in\CC$.  
Thus in the first column and row of the matrix $\rho(g)$ 
we have 
\beqa
\rho(g)_{\lambda\lambda_1}&=&
c\cdot\rho\left
(\prod_{\alpha\in\Phi^+-{\Phi'}^+}(1+c_{-\alpha} e_{-\alpha}) 
\right)_{\lambda\lambda_1} \in\ZZ,\n 
\rho(g)_{\lambda_1\lambda}&=&
c\cdot\rho\left
(\prod_{\alpha\in\Phi^+-{\Phi'}^+}(1+c_\alpha e_\alpha)
\right)_{\lambda_1\lambda} \in\ZZ.
\eeqa 
In particular, if $\lambda_1-\lambda=\alpha\in\Phi^+-{\Phi'}^+$, 
we have
\beqa
\rho(g)_{\lambda\lambda_1}&=&
c c_{-\alpha} \in\ZZ,\n
\rho(g)_{\lambda_1\lambda}&=&
c c_\alpha \in\ZZ,\label{1stcolumnandrow}\\
\rho(g)_{\lambda_1\lambda_1}&=&
c  \in\ZZ.\nonumber
\eeqa 
Note, again, that the factors in the multiple 
products are ordered 
according to some fixed ordering of the roots. If $\lambda$ is a 
weight but cannot be reached from the highest weight $\lambda_1$
by a single Weyl reflection, then $\rho(g)_{\lambda\lambda_1}$ 
($\rho(g)_{\lambda_1\lambda}$) is expressed as a polynomial of 
$c_{-\alpha}$ ($c_\alpha$).

We have defined $E_{\ZZsub}$ to be the group generated by 
$\{\exp e_\alpha | \alpha\in\Phi\}$. We will now show that if 
$g\in G(\ZZ)$, then $g$ can be reduced, by multiplying some 
elements 
in $E_{\ZZsub}$ to $g$, to an element of a subgroup $G'(\ZZ)$  
constructed from $G'\subset G$ with 
$\mbox{rank}\; G'<\mbox{rank}\; G$. 
Once if this is proved, then repeating these operations, 
one may reduce any $g$ to the identity. 
By induction this will show that $G(\ZZ)=E_{\ZZsub}$.

Let us first consider the case when $G$ is simply laced. 
In this case
we fix a lexicographic order in $\Phi^+$ defined by the 
inner product 
in the weight space with some ordered basis $\mu_i$, 
$i=1,\ldots,\mbox{rank}\; G$. (The reason for why we need 
this will be
explained soon.) For example, one may first 
define a partial order in $\Phi^+$ by the height, that is, the 
inner product $\mu_1=\delta$, half the sum of positive 
roots. For 
$\alpha,\beta\in\Phi^+$ with the same height, one may next take 
some arbitrary $\mu$ (linearly independent of $\delta$) and 
define 
$\alpha<\beta$ if $(\mu,\alpha)<(\mu,\beta)$. For those 
which have 
the same inner product with $\mu$, one may consider the product 
with another independent $\mu'$. In this way one may have a 
lexicographic order in $\Phi^+$\footnote{This lexicographic 
order 
can be different from the order introduced above equation 
(\ref{Birkhoff}), 
but they can be made consistent if $\mu_1$ is 
taken $=\lambda_1$. 
Thus we use the same symbol $<$ here for simplicity.
}.

Suppose first that $\rho(g)_{\lambda\lambda_1}\neq 0$, 
$\lambda=\lambda_1-\alpha$ for the minimal 
$\alpha\in\Phi^+-{\Phi'}^+$ 
with respect to the above lexicographic order
(that is, $\alpha < \beta$ for any $\beta\neq\alpha$ 
in $\Phi^+-{\Phi'}^+$).
Let us put $t_\alpha=\exp e_\alpha\in E_{\ZZsub}$, and 
$s_\alpha= t_\alpha^{-1} t_{-\alpha} t_\alpha^{-1}\in 
E_{\ZZsub}$ 
as before. Then for $n\in\ZZ$ we have  
\beqa
\rho(t_{-\alpha}^n s_\alpha g)_{\lambda\lambda_1}
&=&n\rho(g)_{\lambda_1\lambda_1}-\rho(g)_{\lambda\lambda_1},\n
\rho(t_{-\alpha}^n s_\alpha g)_{\lambda_1\lambda_1}
&=&-\rho(g)_{\lambda\lambda_1}.
\eeqa
There exists some integer $n$ such that 
$\rho(t_{-\alpha}^n s_\alpha g)_{\lambda\lambda_1}<
\rho(g)_{\lambda\lambda_1}$. Therefore, by repeating 
this operation, 
one can reduce the $(\lambda,\lambda_1)$ entry of $\rho(g)$ to 0 
(Euclidean algorithm).

We next assume that there exists some $\beta\in\Phi^+-{\Phi'}^+$ 
such that $\rho(g)_{\lambda\lambda_1}=0$ for all 
$\lambda=\lambda_1-\alpha$, $\alpha<\beta$, and 
$\rho(g)_{\lambda_1-\beta,\lambda_1}\neq0$. 
Then applying a similar operation to $\lambda_1-\beta$, we have 
$\rho(u_-g)_{\lambda_1-\beta,\lambda_1}=0$ for some 
element $u_-$ of 
$E_{\ZZsub}$.

Thus, by induction, we have 
$\rho(u_-g)_{\lambda_1-\alpha,\lambda_1}=0$
for all $\alpha\in\Phi^+-{\Phi'}^+$ for some element 
$u_-\in E_{\ZZsub}$.

Let $\alpha\in\Phi^+-{\Phi'}^+$, that is, 
a positive root that contains 
$\alpha_1$. Then $\rho(e_{-\alpha})$ does not 
kill $v_{\lambda_1}$: 
$\rho(e_{-\alpha})v_{\lambda_1}\neq 0$. For 
any minimal representation 
of a simply laced simple Lie group $G$, such $\alpha$ 
can contain only one 
$\alpha_1$. Therefore, any root in $\Phi^+-{\Phi'}^+$ 
cannot be a sum
of any other two such roots. 
Then the order in $\Phi^+-{\Phi'}^+$ 
(induced by that in $\Phi^+$) ensures that the 
inductive assumption 
$\rho(g)_{\lambda\lambda_1}=0$ for all 
$\lambda=\lambda_1-\alpha$, 
$\alpha<\beta$ can never be violated by the operation for 
$\lambda_1-\beta$.

We may now write $u_-g$ in the Birkhoff decomposition as 
\beqa
u_-g&=& \prod_{\alpha\in{\Phi'}^+}\exp(c'_{-\alpha} 
e_{-\alpha})\cdot
w(g) \cdot 
\prod_{\alpha\in\Phi^+}\exp(c'_\alpha e_\alpha).
\label{ug}
\eeqa
Therefore, for any weight $\lambda\neq\lambda_1$, we have 
$\rho(u_-g)_{\lambda\lambda_1}=0$ and 
$\rho(u_-g)_{\lambda_1\lambda_1}=\pm 1$ for some 
$u\in E_{\ZZsub}$.       
Actually we may take $\rho(u_-g)_{\lambda_1\lambda_1}=+1$ since 
$s_\alpha^2=-1$. Using (\ref{1stcolumnandrow}), we find 
\beqa
\rho(u_-g)_{\lambda\lambda_1}&=&
0,\n
\rho(u_-g)_{\lambda_1\lambda}&=&
c'_\alpha \in\ZZ,\label{1stcolumnandrowofu-g}\\
\rho(u_-g)_{\lambda_1\lambda_1}&=&
1,\nonumber
\eeqa 
where $\alpha\in\Phi^+-{\Phi'}^+$.
Thus $u_+=\prod_{\alpha\in\Phi^+-{\Phi'}^+}
\exp(-c'_\alpha e_\alpha)$
(in an arbitrary order) belongs to $E_{\ZZsub}$, and 
\beqa
\rho(u_-gu_+)_{\lambda\lambda_1}&=&
0,\n
\rho(u_-gu_+)_{\lambda_1\lambda}&=&
0,\label{1stcolumnandrowofu-gu+}\\
\rho(u_-gu_+)_{\lambda_1\lambda_1}&=&
1\nonumber
\eeqa 
for all the weights $\lambda\neq\lambda_1$. Therefore, $u_-gu_+$ 
belongs to the subgroup $G'\subset G$ of whose simple 
roots are $\Delta-\alpha_1$, and the minimal 
representation $\rho$ 
of $G$ is again decomposed into a direct sum of minimal 
representations 
of $G'$. This is what we desired, and completes 
the proof for a simply laced Lie group $G$.

Finally, if $G$ is not simply laced, the minimal 
representation $\rho$ 
is either the ${\bf 2}^{\mbox{\boldmath $N$}}$ for $SO(2N+1)$ or 
the ${\bf 2}\mbox{\boldmath $N$}$ for $Sp(2N)$. One may 
also apply 
the same argument to these cases except for the modification 
in the 
consistent order of operations; since some roots in 
$\Phi^+-{\Phi'}^+$ 
can be written as a sum of two roots in $\Phi^+-{\Phi'}^+$ if  
$G$ is not simply laced, the lexicographic order fails 
to ensure the 
consistency of induction. Nevertheless, one may still find 
alternative consistent orders of operations explicitly, and 
prove the proposition in both cases. We omit the detail in this 
paper; a way to find this is to write the Hasse diagrams for 
these representations. 

\newcommand{\NP}[1]{Nucl.\ Phys.\ {\bf #1}}
\newcommand{\PL}[1]{Phys.\ Lett.\ {\bf #1}}
\newcommand{\CMP}[1]{Comm.\ Math.\ Phys.\ {\bf #1}}
\newcommand{\PR}[1]{Phys.\ Rev.\ {\bf #1}}
\newcommand{\PRL}[1]{Phys.\ Rev.\ Lett.\ {\bf #1}}
\newcommand{\MPL}[1]{Mod.\ Phys.\ Lett.\ {\bf #1}}

\end{document}